\documentclass[sigconf]{acmart}
\copyrightyear{2021}
\acmYear{2021}
\setcopyright{acmlicensed}\acmConference[CCS '21]{Proceedings of the 2021 ACM SIGSAC Conference on Computer and Communications Security}{November 15--19, 2021}{Virtual Event, Republic of Korea}
\acmBooktitle{Proceedings of the 2021 ACM SIGSAC Conference on Computer and Communications Security (CCS '21), November 15--19, 2021, Virtual Event, Republic of Korea}
\acmPrice{15.00}
\acmDOI{10.1145/3460120.3484583}
\acmISBN{978-1-4503-8454-4/21/11}

\usepackage{xspace}
\usepackage{array,multirow,graphicx}
\usepackage{float}
\usepackage{adjustbox}
\usepackage[utf8]{inputenc}
\usepackage{tcolorbox}
\usepackage{enumitem} 

\makeatletter
\renewcommand\paragraph{\@startsection{paragraph}{4}{\parindent}%
  {-.25\baselineskip \@plus -2\p@ \@minus -.2\p@}
  {-3.5\p@}%
  {\ACM@NRadjust{\@parfont\@adddotafter}}}
\makeatother

\newif\ifshowcomments
\showcommentstrue

\ifshowcomments
\newcommand{\mybetternote}[2]{\xspace\fbox{\bfseries\sffamily\scriptsize{#1}}
	{\small$\blacktriangleright$\textsf{\emph{#2}}$\blacktriangleleft$}}
\else
\newcommand{\mybetternote}[2]{}
\fi

\newcommand{\leo}[1]{\textcolor{blue}{\mybetternote{LQ}{#1}}}
\newcommand{\dan}[1]{\textcolor{orange}{\mybetternote{DCDE}{#1}}} 
\newcommand{\pie}[1]{\textcolor{cyan}{\mybetternote{PB}{#1}}} 

\newcommand{\edit}[1]{{\color{black}#1}}
\newcommand{\scz}[1]{{\color{black}#1}}
\newcommand{\cready}[1]{{\color{black}#1}}
\newcommand{\opensourcerepo}{{\color{black}\url{https://github.com/pietroborrello/constantine}}}

\newcommand{\proj}{\textsc{Constantine}\xspace}
\newcommand{\taken}{\textit{taken}\xspace}

\usepackage{subcaption}
\usepackage{listings}
\lstdefinestyle{myVerbatimStyle} {
    basicstyle=\ttfamily,
    mathescape=true
}
\lstdefinestyle{myVerbatimSmaller} {
    basicstyle=\ttfamily\footnotesize,
    mathescape=true
}
\lstdefinestyle{inlineCode}{
    basicstyle=\ttfamily\footnotesize,
    commentstyle=\fontseries{lc}\selectfont\itshape,
    columns=fullflexible
}
\usepackage[normalem]{ulem}
\newcommand{\stkout}[1]{\ifmmode\text{\sout{\ensuremath{#1}}}\else\sout{#1}\fi}
\begin{document}
\fancyhead{}

\title[{\sc Constantine}: Automatic Side-Channel Resistance\\Using Efficient Control and Data Flow Linearization]{{\sc Constantine}: Automatic Side-Channel Resistance Using Efficient Control and Data Flow Linearization}

\author{Pietro Borrello}
\affiliation{\institution{\normalsize Sapienza University of Rome}}
\email{borrello@diag.uniroma1.it}

\author{Daniele Cono D'Elia}
\affiliation{\institution{\normalsize Sapienza University of Rome}}
\email{delia@diag.uniroma1.it}

\author{Leonardo Querzoni}
\affiliation{\institution{\normalsize Sapienza University of Rome}}
\email{querzoni@diag.uniroma1.it}

\author{Cristiano Giuffrida}
\affiliation{\institution{\normalsize Vrije Universiteit Amsterdam}}
\email{giuffrida@cs.vu.nl}


\begin{abstract}
\cready{In the era of microarchitectural side channels, vendors scramble to deploy mitigations for transient execution attacks, but leave traditional side-channel attacks against sensitive software (e.g., crypto programs) to be fixed by developers by means of \emph{constant-time programming} (i.e., absence of secret-dependent code/data patterns)}. Unfortunately, writing constant-time code by hand is hard, as evidenced by the many flaws discovered in production side channel-resistant code. Prior efforts to automatically transform programs into constant-time equivalents offer limited security or compatibility guarantees, hindering their applicability to real-world software.

In this paper, we present \proj, a compiler-based system to automatically harden programs against microarchitectural side channels. \proj pursues a radical design point where secret-dependent control and data flows are \emph{completely linearized} (i.e., all involved code/data accesses are always executed). This strategy provides strong security and compatibility guarantees by construction, but its natural implementation leads to state explosion in real-world programs. To address this challenge, \proj relies on carefully designed optimizations such as just-in-time loop linearization and aggressive function cloning for fully context-sensitive points-to analysis, which not only address state explosion, but also lead to an efficient and compatible solution. \proj yields overheads as low as \cready{$16\%$} on standard benchmarks and can handle a fully-fledged component from the production wolfSSL library.
\end{abstract}



\maketitle


\section{Introduction}
Protecting the confidentiality of security-sensitive information is a key requirement of modern computer systems.
Yet, despite advances in software security engineering, this requirement is more and more challenging to satisfy in face of increasingly sophisticated microarchitectural side-channel attacks. Such attacks allow adversaries to leak information from victim execution by observing changes in the microarchitectural state (e.g., cache eviction), typically via timing measurements (e.g., memory access latency).

\cready{Such attacks have been shown practical in the real world with or without the assistance of CPU bugs. Examples in the former category are transient execution attacks such as Spectre~\cite{kocher-sp19}, Meltdown~\cite{lipp-usenix18}, L1TF~\cite{bulk-usenix18}, and MDS~\cite{VanSchaik-sp19,Schwarz-ccs19,Canella-ccs19}. Examples in the latter category are traditional cache attacks (e.g., FLUSH+RELOAD~\cite{yarom-usenix14} and PRIME+PROBE~\cite{osvik-rsa06}) against security-sensitive software victims such as crypto libraries. While the former are the focus of many mitigation efforts by vendors, for the latter the burden of mitigation lies entirely on the shoulders of software developers~\cite{intelstmt}.}

In theory, this is feasible, as side-channel attacks leak secrets (e.g., crypto keys) by observing victim secret-dependent computations (e.g., branch taken or array indexed based on a crypto key bit) via microarchitectural measurements. Hence, eliminating explicit secret-dependent code/data accesses from software---a practice generally known as \emph{constant-time programming}~\cite{Barthe18}---is a viable avenue for mitigation. In practice, removing side-channel vulnerabilities from software is a daunting and error-prone task even for skilled developers. Not surprisingly, even production side channel-resistant implementations are riddled with flaws~\cite{Somorovsky-CCS16,binsecrel-SP20}.

To address this problem, much prior work has proposed solutions to automatically transform programs into their constant-time equivalents or variations~\cite{GHOSTRIDER-ASPLOS15,MTO-CSF13,PHANTOM-CCS13,timewarp-isca12,Molnar-ICISC05,ASCEND-STC12,fletchery-hpca2014,stealthmem-usenix12,Zhang-PLDI12,Shi-asiacrypt2011,PATHORAM-CCS13,aegis-sc2003,hunger-hpca2015,vattikonda-wccs2011,wang-isca2007,wang-microarchitecture2008,Zhang-CCS11,zhang-sp2011,zhang-ccs2013,FaCT-PLDI19}. Unfortunately, even the most recent solutions~\cite{RACCOON-SEC15,Wu-ISSTA18,Soares-CGO21} offer limited security or compatibility guarantees, hindering their applicability to real-world programs.

In this paper, we introduce \proj, a compiler-based system for the automatic elimination of side-channel vulnerabilities from programs. The key idea is to explore a radical design point based on \emph{control and data flow linearization} (or CFL and DFL), where all the possible secret-dependent code/data memory accesses are always executed regardless of the particular secret value encountered. The advantage of this strategy is to provide strong security and compatibility guarantees by construction. The nontrivial challenge is to develop this strategy in a practical way, since a straightforward implementation would lead to \emph{program state explosion}. For instance, naively linearizing secret-dependent branches that guard loop exits would lead to unbounded loop execution. Similarly, naively linearizing secret-dependent data accesses by touching all the possible memory locations would lead to an unscalable solution.

Our design is indeed inspired by radical and impractical-by-design obfuscation techniques such as the M/o/Vfuscator~\cite{movfuscator}, which linearizes the control flow to collapse the program's control-flow graph into a single branchless code block with only data movement (i.e., x86 {\tt mov}) instructions~\cite{Kirsch-IFIPSEC17}. Each {\tt mov} instruction uses an extra level of indirection to operate on real or dummy data depending on whether the code is running the intended side of a branch or not.

Revisiting such design point for side-channel protection faces several challenges. First, linearizing all the branches with {\tt mov}-only code hinders efficient code generation in modern compilers and leads to enormous overheads. To address this challenge, \proj only linearizes secret-dependent branches pinpointed by profiling information, allows arbitrary branchless instructions besides {\tt mov}, and uses efficient indirect memory addressing to allow the compiler to generate efficient code. Second, the M/o/Vfuscator only linearizes the control flow and encodes branch decisions in new data flows, a strategy which would only multiply the number of secret-dependent data accesses. To address this challenge, \proj couples CFL with DFL to also linearize all the secret-dependent data flows (generated by CFL or part of the original program).

Finally and above all, M/o/Vfuscator does not address state explosion. For example, it linearizes loop exits by means of invalid {\tt mov} instructions, which generate exceptions and restart the program in dummy mode until the original loop code is reached. Other than being inefficient, this strategy originates new side channels (e.g., exception handling) that leak the number of loop iterations. To address state explosion, \proj relies on carefully designed optimizations such as \emph{just-in-time loop linearization} and \emph{aggressive function cloning}. The former linearizes loops in the same way as regular branches, but adaptively bounds the number of iterations based on the original program behavior. The latter enables precise, context-sensitive points-to analysis which can strictly bound the number of possible targets at secret-dependent data accesses.

Collectively, our optimizations produce a scalable CFL and DFL strategy, while supporting all the common programming constructs in real-world software such as nested loops, indirect function calls, pointer-based accesses, etc. Our design not only addresses the state explosion problem, but also leads to a system that outperforms prior comprehensive solutions in terms of both performance and compatibility, while also providing stronger security guarantees. For example, we show \proj yields overheads as low as \cready{$16\%$} for cache-line attacks on standard benchmarks. Moreover, to show \proj provides the first practical solution for automatic side-channel resistance for real-world software, we present a case study on the wolfSSL embedded TLS library. We show \proj{}-protected wolfSSL can complete a modular multiplication of a ECDSA signature in 8~ms, which demonstrates \proj's automated approach can effectively handle a fully-fledged real-word crypto library component for the very first time.

\subsubsection*{Contributions}
To summarize, this paper proposes the following contributions:
\begin{itemize}
	\item We introduce \proj, a system for the protection of software from side channels.
	\item We show how \proj can automatically analyze and transform a target program by efficiently applying control and data flow linearization techniques.
	\item We implement \proj as a set of compiler transformations for the LLVM toolchain. \proj is open source \cready{(available at \opensourcerepo)}.
	\item We evaluate \proj on several standard benchmarks, evidencing its performance advantage against prior solutions. We also present a case study on the wolfSSL library to show its practical applicability on real-world software.
\end{itemize}

\vspace{-2mm}
\section{Background}
\label{se:background}

Microarchitectural side channels generally allow an adversary to infer \emph{when} and \emph{where} in memory a victim program performs specific code/data accesses. And by targeting secret-dependent accesses originating from secret-dependent control and data flows in the original program, an adversary can ultimately leak secret data. Constant-time programming is a promising solution to eliminate such explicit secret-dependent accesses from programs, but state-of-the-art automated solutions are severely limited in terms of security, performance, and/or compatibility. 

%

\vspace{-0.29em}
\paragraph{Control Flow}
Secret-dependent control flows (e.g., code branching on a crypto key bit) induce code accesses that microarchitectural attacks can observe to leak secrets. Early constant-time programming solutions only attempted to balance out secret-dependent branches with dummy instructions (e.g., with cross-copying~\cite{Agat-POPL00}) to mitigate only simple execution time side-channel attacks~\cite{Mantel-ESORICS15}. Molnar et al.~\cite{Molnar-ICISC05} made a leap forward with the \textit{program counter security model} (PC-security), where the trace of secret-dependent executed instructions is the same for any secret value.

Prior work has explored two main avenues to PC-security. The first avenue is a form of \textit{transactional execution}~\cite{RACCOON-SEC15}, which always executes both sides of every secret-dependent branch---hence a \textit{real} and a \textit{decoy} path---as-is, but uses a transaction-like mechanism to buffer and later discard changes to the program state from decoy paths. This approach provides limited security guarantees, as it introduces new side channels to observe decoy path execution and thus the secret. Indeed, one needs to at least mask exceptions from rogue operands of read/write instructions on decoy paths, introducing secret-dependent timing differences due to exception handling. Even when normalizing such differences, decoy paths may perform read/write accesses that real paths would not make, introducing new decoy data flows. An attacker can easily learn data-flow invariants on real paths (e.g., an array always accessed at the same offset range) and detect decoy path execution when the observed accesses reveal invariant violations. See Appendix~\ref{apx:example} for concrete decoy path side channel examples. Also, this approach alone struggles with real-world software compatibility. For instance, it requires loops to be completely unrolled, which leads to code size explosion for
nested loops and for those with large trip count.

Another avenue to PC-security is \textit{predicated execution}~\cite{Coppens-SP09}, which similarly executes both real and decoy paths, but only allows the instructions from the real path to update the program state. Updates are controlled by a predicate that reflects the original program branch condition and take the form of a constant-time conditional assignment instruction (e.g., {\tt cmov} on x86)~\cite{Coppens-SP09}. When on a decoy path, read/write operations get rewired to a single (conditionally assigned) shadow address. However, such decoy (shadow) data flows can again introduce new side channels to leak the decoy nature of a path~\cite{Coppens-SP09, vanCleemput-TACO12}. Moreover, this form of predication hampers the~\pagebreak[4] optimization process of the compiler, forcing the use of pervasive {\tt cmov} instructions and constraining code transformation and generation. Some more recent solutions attempt to generate more optimized code by allowing some~\cite{Soares-CGO21} or all~\cite{Wu-ISSTA18} accesses on unmodified addresses on decoy paths. However, this hybrid strategy mimics transactional execution behavior and is similarly vulnerable to side-channel attacks that detect data-flow invariant violations on decoy paths. In addition, existing solutions face the same compatibility issues of transactional solutions with real-world code.

Unlike prior solutions, \proj's control-flow linearization (CFL) executes both real and decoy paths using an indirect memory addressing scheme to transparently target a shadow address along decoy paths. This strategy does not force the compiler to use {\tt cmov} instructions and yields more efficient generated code. For instance, a \proj{}-instrumented wolfSSL binary contains only \text{39}\% of {\tt cmov} instructions (automatically emitted by the code generator, as appropriate) compared to predicated execution, resulting in a net CFL speedup of \text{32.9}\%. As shown later, the addition of data-flow linearization (DFL) allows \proj to operate further optimizations, eliminating shadow address accesses altogether as well as the corresponding decoy data flows. \proj is also compatible with all the common features in real-world programs, including variable-length loops bound by means of \emph{just-in-time linearization} and indirect calls handled in tandem with DFL.

\vspace{-0.29em} 
\paragraph{Data Flow}
Data-dependent side channels have two leading causes on modern microarchitectures. Some originate from instructions that exhibit data operand-dependent latencies, e.g., integer division~\cite{Coppens-SP09} and some floating-point instructions~\cite{Andrysco-SP15} on x86. Simple mitigations suffice here, including software emulation~\cite{CFTP-CCS18} (also adopted by \proj), compensation code insertion~\cite{vanCleemput-TACO12}, and leveraging hardware features to control latencies~\cite{Coppens-SP09}.

The other more fundamental cause stems from secret-dependent data flows (e.g., an array accessed at an offset based on a crypto key bit), which induce data accesses that microarchitectural attacks can observe to leak secrets. 
Hardware-based mitigations~\cite{Wang-ISCA07} do not readily apply to code running on commodity processors. To cope with such leaks, existing compiler-based solutions have explored code transformations~\cite{Wu-ISSTA18} and software-based ORAM~\cite{RACCOON-SEC15}. 

SC-Eliminator~\cite{Wu-ISSTA18} transforms code to preload cache lines of security-sensitive lookup tables, so that subsequent lookup operations can result in always-hit cache accesses, and no secret-dependent time variance occurs. Unfortunately, since the preloading and the secret-dependent accesses are not atomic, a non-passive adversary may evict victim cache lines right after preloading and later observe the secret-dependent accesses with a cache attack. Cloak~\cite{gruss-usenix17} adopts a similar mitigation approach, but enforces atomicity by means of Intel TSX transactions. Nonetheless, \cready{it} requires manual code annotations and can only support short-lived computations. Moreover, these strategies are limited to standard cache attacks and do not consider other microarchitectural attacks, including those that operate at the subcacheline granularity~\cite{CacheBleed-CHES16,MemJam}.

Raccoon~\cite{RACCOON-SEC15} uses Path ORAM (Oblivious RAM)~\cite{PATHORAM-CCS13} as a shortcut to protect data flows from attacks. ORAMs let code conceal its data access patterns by reshuffling contents and accessing multiple cells at each retrieval~\cite{ORAM-JACM96}. Unfortunately, this strategy introduces substantial run-time overhead as each security-sensitive data access results in numerous ORAM-induced memory accesses.\newpage

Unlike prior solutions, \proj's data-flow linearization (DFL) eliminates all the explicit secret-dependent data flows (generated by CFL or part of the original program) by forcing the corresponding read/write operations to touch all their target memory locations as computed by static points-to analysis. While such analyses are known to largely overapproximate target sets on real-world programs, \proj relies on an \emph{aggressive function cloning strategy} to enable precise, context-sensitive points-to analysis and strictly bound the number of possible targets. For instance, a \proj{}-instrumented wolfSSL binary using state-of-the-art points-to analysis~\cite{SVF-CC16} yields an average number of target objects at secret-dependent data accesses of $\text{6.29}$ and $\text{1.08}$ before and after aggressive cloning (respectively), a net reduction of $\text{83}\%$ resulting in precise and efficient DFL. Unlike prior solutions that are limited to array accesses, \proj is also compatible with arbitrary pointer usage in real-world programs.

\section{Threat Model}
\label{se:threatmodel}
\cready{We assume a strong adversary able to run arbitrary code on the target machine alongside the victim program, including on the same physical or logical core. The adversary has access to the source/binary of the program and seeks to leak secret-dependent computations via microarchitectural side channels. Other classes of side channels (e.g., power~\cite{Lipp2021Platypus}), software (e.g., external libraries/OS~\cite{bosman2016dedup}) or hardware (e.g., transient execution~\cite{Cauligi-PLDI20,Schwarzl2021Specfuscator}) victims, and vulnerabilities (e.g., memory errors or other undefined behaviors~\cite{wang2013towards}) are beyond the scope of constant-time programming and subject of orthogonal mitigations}. We further make no restrictions on the microarchitectural side-channel attacks attempted by the adversary, ranging from classic cache attacks~\cite{yarom-usenix14,osvik-rsa06} to recent contention-based attacks~\cite{Aldaya-sp19,gras-ndss20}. With such attacks, we assume the adversary can observe the timing of arbitrary victim code/data accesses and their location at the cache line or even lower granularity. 

\section{\proj}
\label{se:constantine}
This section details the design and implementation of \proj. We first outline its high-level workflow and building blocks.

\subsection{Overview}
\proj is a compiler-based system to automatically harden programs against microarchitectural side channels. Our linearization design pushes constant-time programming to the extreme and embodies it in two flavors:

\begin{enumerate}
\item \textbf{Control Flow Linearization} (CFL): we transform program regions influenced by secret data to yield secret-invariant instruction traces, with real and decoy parts controlled by a \textit{dummy execution} abstraction opaque to the attacker;
\item \textbf{Data Flow Linearization} (DFL): we transform every secret-dependent data access (including those performed by dummy execution) into an oblivious operation that touches all the locations such program point can possibly reference, leaving the attacker unable to guess the intended target.
\end{enumerate}



CFL and DFL add a level of {\em indirection} around value computations. The CFL dummy execution abstraction uses it to implicitly nullify the effects of instructions that presently execute as decoy paths. DFL instead wraps load and store operations to induce memory accesses for the program that are secret-invariant, also ensuring real and decoy paths access the same collections of objects.


Linearizing control and data flows represents a radical design point with obvious scalability challenges. To address them, \proj relies on carefully designed optimizations. For control flows, we rely on a M/o/Vfuscator-inspired \emph{indirect memory addressing} scheme to legalize decoy paths while allowing the optimizer to see through our construction and generate efficient code. We also propose \textit{just-in-time loop linearization} to efficiently support arbitrary loops in real-world programs and automatically bound their execution based on the behavior of the original program (i.e., automatically padding the number of iterations based on the maximum value observed on real paths).

For data flows, we devise \textit{aggressive function cloning} to substantially boost the precision of static memory access analysis and minimize the extra accesses required by DFL. To further optimize DFL, we rely on an efficient object metadata management scheme and on hardware-optimized code sequences (e.g., AVX-512) to efficiently touch all the necessary memory locations at each secret-dependent data access. We also exploit synergies between control-flow and data-flow handling to (i)~eliminate the need for shadow accesses on decoy paths (boosting performance and eradicating problematic decoy data flows altogether); (ii)~handle challenging indirect control flows such as indirect function calls in real-world programs.


To automatically identify secret-dependent code and data accesses, we rely on \textit{profiling} information obtained via dynamic information flow tracking and propagate the dependencies along the call graph. \cready{To analyze memory accesses}, we consider a state-of-the-art Andersen-style points-to analysis implementation~\cite{SVF-CC16} and show how aggressive function cloning can greatly boost its precision thanks to newly added full context sensitivity.
%
%

From a security perspective, CFL ensures PC-security for all the instructions that operate on secret data or whose execution depends on it; in the process it also replaces variable-latency instructions with safe software implementations. DFL provides analogous guarantees for data: at each secret-dependent load or store operation, the transformed program obliviously accesses every potentially referenced location in the execution for that program point and is no longer susceptible to microarchitectural leaks by design.

\begin{figure*}[!t]
\vspace{-0.35em}
\centering
\includegraphics[width=0.85\textwidth]{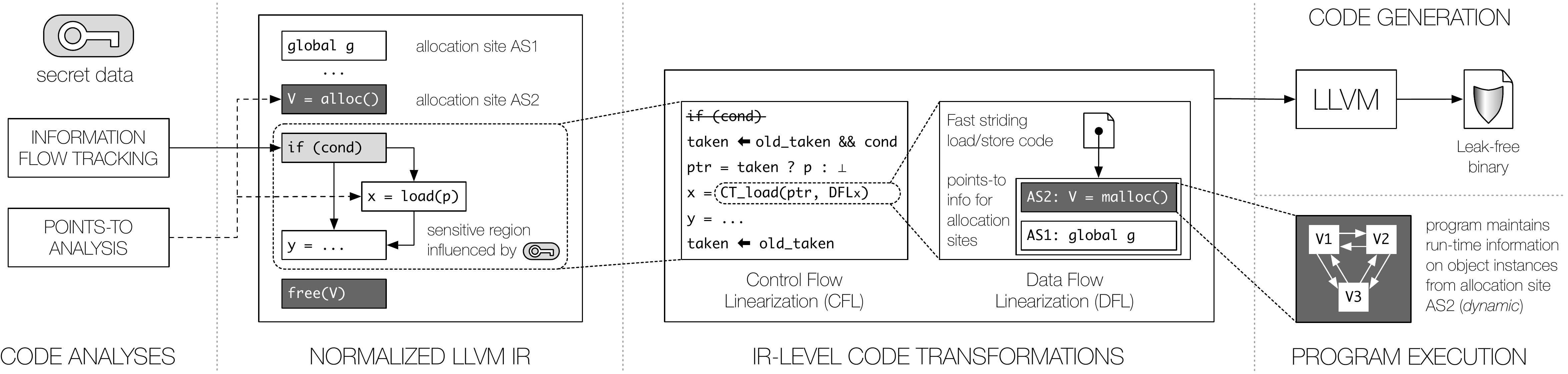}
\vspace{-1em}
\caption{Architecture of \proj: code analyses, CFL \& DFL transformations, and run-time object metadata\label{fig:architecture}.}
\vspace{-1mm}
\end{figure*}

Figure~\ref{fig:architecture} provides a high-level view of the CFL, DFL, and support program analysis components behind \proj. Our techniques are general and we implement them as analyses and transformation passes for the \textit{intermediate representation} (IR) of LLVM.



\vspace{-1pt}
\subsection{Control Flow Linearization}
\label{se:cfl}

With control flow linearization (CFL) we turn \edit{secret-dependent control flows into straight-line regions} 
that meet PC-security requirements by construction~\cite{Molnar-ICISC05}, \edit{proposing just-in-time linearization for looping sequences}. We also make provisions for instructions that may throw an exception because of rogue values along decoy paths, or yield variable latencies because of operand values. 


{\centering
\begin{tcolorbox}[fontupper=\itshape\textsf, width=0.975\columnwidth,boxsep=0pt,top=6pt, bottom=6pt, left=10pt, right=10pt]
\small
\textbf{CFL}: The sequence of secret-dependent instructions that the CPU executes is constant for any initial input (PC-security) and data values do not affect the latency of each such instruction. 
\end{tcolorbox}
\par}

\scz{With this invariant,} only data access \edit{patterns} can then influence execution time, and DFL will make them insensitive to secret input values. We assume that an oracle (the taint analysis of \textsection\ref{ss:taint}) enucleates which control-flow transfer decisions depend on secret data. Such information comprises if-else and loop constructs and indirect-call targets. For each involved code region, we push the linearization process in a recursive fashion to any nested control flows (i.e., if-else branches, loops, and function calls), visiting control-flow graphs (CFGs) and call graph edges in a post-order depth-first fashion. \edit{By doing so we avoid leaks from decoy paths when executing secret-independent inner branches in a protected region.}

\subsubsection{Dummy Execution}
\label{ss:dummy}
Each linearized region holds a \textit{``taken''} predicate instance that determines if the original program would execute it (\textit{real path}) or not (\textit{decoy path}) under the current program state. We incrementally update the predicate \edit{with a new instance} at every control-flow decision that guards the region in the original program, and \edit{let the compiler use the previous incoming instance} upon leaving the region.
The predicate backs a {\em dummy execution} indirection abstraction where we let decoy paths execute together with real paths, and use the \taken predicate to prevent that visible effects from decoy paths may pollute the program state. 

The key to correctness is that we can safely allow decoy paths to make local computations (i.e., assign to virtual registers in the IR), as long as their values do not flow into memory.
For memory operations, each pointer expression computation selects an artificial $\bot$ value when in dummy execution. DFL primitives wrap every load and store instruction and make both real and decoy paths stride the same objects thanks to points-to metadata associated with the memory operation. Upon leaving a region, local values that the program may use later (i.e. \textit{live} virtual registers) undergo a selection step to pick values from real paths at merge points.



The key to efficiency is using a selection primitive that is transparent for the optimizer thanks to indirection. As we observed in \textsection\ref{se:background}, the {\tt cmov} selector typical of predicated execution constrains the behavior of \cready{the optimizer during code generation. We leverage the indirection on \taken to design selection primitives based on arithmetic and logic operations that can instead favor optimizations.

Let us consider the pointer assignment $ptr = taken\mathrel{?}p\mathrel{:}\bot$ of Figure~\ref{fig:architecture}. By modeling \taken as an integer being $1$ on real paths and $0$ on decoy ones, and by using NULL to represent $\bot$ for DFL, the selection becomes $ptr = taken\mathrel{*}p$. DFL helpers will prevent NULL accesses and deem them as from decoy paths: those cannot happen on real paths since, like prior literature~\cite{RACCOON-SEC15}, we work on error-free programs. This constant-time multiplication-based scheme unleashes many arithmetic optimizations (e.g., global value numbering~\cite{Rosen-POPL88}, peephole~\cite{ullman06}) at the IR and backend level, bringing a net CFL speedup of \text{32.9}\% in wolfSSL over using the {\tt cmov} approach. Appendix~\ref{apx:select} details other primitives that we evaluated. 

Selection may be needed for ($\phi$) compiler temporaries too, as we will detail in \textsection\ref{ss:branches}. Unlike memory addresses, both incoming values may be arbitrary, allowing for more limited optimization: for them we use the {\tt select} IR instruction and let LLVM lower it branchlessly as it sees fit (including an x86 {\tt cmov}).

Hereafter, we use {\tt ct\_select} to refer to a constant-time selection of any values, but we inline the logic in the IR in the implementation.}


\subsubsection{Compiler IR Normalization}
\label{ss:ir-norm}
\proj takes as input the intermediate representation (IR) produced for the program by the language-specific compiler frontend. We assume that the IR comes in static single assignment (SSA) form~\cite{Rosen-POPL88} and that the CFG of every function containing regions to transform is reducible.
\edit{The code can come in already-optimized form (e.g., {\tt -O2}, {\tt -O3} settings)}.

We apply a number of {\em normalization} passes that simplify later transformations with the ultimate goal of having {\em single-entry, single-exit regions} as unit of transformation, similarly to~\cite{Wu-ISSTA18}.

We use existing LLVM passes to lower switch constructs into if-else sequences, and to unify multiple function exit points into a single one (for abort-like sequences that do not fall through, we add artificial CFG edges to the exit node). {\color{black} As we work on error-free programs, we replace exception-aware {\tt invoke} statements with normal calls. We also turn indirect calls into if-else sequences of direct calls using \cready{points-to} information (\textsection\ref{ss:points-to}), \edit{guarding each direct call with a pointer comparison on the target}.



We then massage the CFG using standard compiler techniques~\cite{ullman06} so that it results into a graph composed only of single-entry, single-exit regions: this will hold for all branches and loop constructs in the IR. This normalized IR is the input for the taint oracle of \textsection\ref{ss:taint}.



\subsubsection{Branch Linearization}
\label{ss:branches}
We can now detail how branch linearization operates and its orchestration with dummy execution. Under the single-entry, single-exit structural assumption from IR normalization, for a conditional construct of the likes \textit{if (cond) then \{A\} else \{B\}}, we note that its exit CFG node post-dominates both the ``then'' and ``else'' regions of the branch, and is dominated by the entry node by construction. \edit{In SSA form, $\phi$-nodes select incoming path-sensitive values.} To linearize a conditional construct we: 
\begin{enumerate}[topsep=0pt]
\item remove the conditional branch, unlinking blocks A and B;
\item replace in A every pointer expression computation with a conditional assignment ${\texttt{ct\_}}\texttt{select}(\textit{cond},\textit{ptr}, \bot)$;
\item replace similarly in B, using the condition negated ($!\textit{cond}$);
\item wrap memory accesses with DFL {\tt ct\_\{load, store\}} primitives, supplying the DFL metadata for the operation (\textsection\ref{se:dfl});
\item replace each $\phi$-node $v_0=\phi(v_A,v_B)$ in the exit block (which assigns virtual register $v_0$ according to whether A or B executed) with a conditional assignment ${\texttt{ct\_}}\texttt{select}(\textit{cond},v_A, v_B)$;
\item merge $\langle$entry, A, B, exit$\rangle$ to form a single block, in this order.
\end{enumerate}

We thus ``sink'' $\textit{cond}$ to conditionally assign pointers ($\bot$ for decoy paths) and virtual registers that outlive the region. Our transformation preserves the SSA form and can always be applied locally.



We can now add the dummy execution idea to the picture. Without loss of generality, let us consider two nested if-else statements that possibly take part in a larger linearized region as in~\pagebreak[4] Figure~\ref{fig:dummy}. When reaching the outer if construct, the program sees a \taken predicate instance $\texttt{t}_0$ that determines whether the \edit{execution} reached the construct \edit{as part of} a real ($\textit{taken}\mathbin{=}\textit{true}$) or decoy computation.

Inside a region, IR instructions that assign virtual registers do not need to know $\texttt{t}_0$. Path-sensitive assignments of live-out values from a region, such as $\texttt{b$_{\texttt{inner}}$}$, check the linearized conditions ($\texttt{c$_{\texttt{inner}}$}$ in this case). Memory-related instructions see instead their pointer expressions conditionally assigned according to some $\texttt{t}_i$ \taken instance. Those instances are updated upon entering the enclosing code block in the (original) program to reflect the combination of control-flow conditions with the incoming \taken predicate.



\begin{figure}[t!]
\begin{scriptsize}
\centering
\begin{subfigure}[b]{0.35\columnwidth}
\centering
\begin{tabular}{c}
\begin{lstlisting}[style=myVerbatimStyle]
if (c$_{\texttt{outer}}$) {
  b$_1$ = v[2]
} else {
  if (c$_{\texttt{inner}}$) {
    b$_2$ = v[0]
  } else {
    b$_{3}$ = 0
  }
  b$_{\texttt{inner}}$ = $\phi$(b$_2$,b$_{3}$)
}
b$_4$ = $\phi$(b$_1$,b$_{\texttt{inner}}$)
v[1] = b$_4$
\end{lstlisting}
\vspace{-1mm}
\end{tabular}
\caption{Original code} 
\end{subfigure}
\begin{subfigure}[b]{0.565\columnwidth}
\centering
\begin{tabular}{c}
\begin{lstlisting}[style=myVerbatimStyle]
t$_0$ = <incoming 'taken' predicate>
t$_1$ = c$_{\texttt{outer}}$ && t$_0$
  ptr$_1$ = ct_select(t$_1$, &v[2], $\bot$)
  b$_1$ = ct_load(ptr$_1$, DFL$_{\texttt{b}_1}$)
t$_{1-{\texttt{else}}}$ = !c$_{\texttt{outer}}$ && t$_0$
  t$_2$ = c$_{\texttt{inner}}$ && t$_{1-{\texttt{else}}}$
    ptr$_2$ = ct_select(t$_2$, &v[0], $\bot$)
    b$_2$ = ct_load(ptr$_2$, DFL$_{\texttt{b}_2}$)
  t$_{2-{\texttt{else}}}$ = !c$_{\texttt{inner}}$ && t$_{1-{\texttt{else}}}$ // unused
    b$_3$ = 0  
  b$_{\texttt{inner}}$ = ct_select(c$_{\texttt{inner}}$, b$_2$, b$_3$)
b$_4$ = ct_select(c$_{\texttt{outer}}$, b$_1$, b$_{\texttt{inner}}$)
ptr$_3$ = ct_select(t$_0$, &v[1], $\bot$)
ct_store(ptr$_3$, b$_4$, DFL$_{\texttt{store}_1}$)
\end{lstlisting}
\vspace{-1mm}
\end{tabular}
\caption{After linearization\label{fig:dummy-addrs}} 
\end{subfigure}
\end{scriptsize}
\vspace{-0.85em}
\caption{Linearization and dummy execution.\label{fig:dummy}}
\vspace{-0.65em}
\end{figure}


\vspace{-1mm}
\subsubsection{Loop Linearization}
\label{ss:loops}

To cope with the practical requirements of real-world code, with \proj we explore a {\em just-in-time} approach for the linearization of loops. Let us consider the following secret-sensitive fragment, taken from a wolfSSL function  that computes $x/R == x~(mod~N)$ using a Montgomery reduction:

\vspace{-1pt}
{\centering
\begin{lstlisting}[language=C, style=myVerbatimSmaller]
  _c   = c + pa;
  tmpm = a->dp;
  for (x = 0; x < pa+1; x++)
    *tmpm++ = *_c++;
  for (; x < oldused; x++)  // zero any excess digits on
    *tmpm++ = 0;  // destination that we didn't write to
\end{lstlisting}
\par}
\vspace{-1pt}

The induction variable $x$ depends on secret data $\textit{pa}$, outlives the first loop, and dictates the trip count of the second loop. Prior solutions struggle with each of these aspects, as well as with continue/break statements we found in wolfSSL. For the secret-dependent trip count issue, some~\cite{Wu-ISSTA18} try to infer a bound and pad the loop with decoy iterations, then unroll the loop completely. However, high trip counts seen at run time or inaccurate bound predictions make unrolling immediately impractical due to code bloat.

In \proj we design a new approach to handle loops that avoids unrolling and supports full expressivity for the construct. The key idea is to flank the normal trip count of a loop with an own CFL induction variable---dubbed $\textit{c\_idx}$ next---and let such variable dictate just-in-time how many times that loop should execute.

\begin{figure}[t!]
\begin{scriptsize}
\centering
\begin{subfigure}[b]{0.42\columnwidth}
\centering
\begin{tabular}{c}
\begin{lstlisting}[style=myVerbatimStyle]
base:
  i$_{\texttt{base}}$ = 0
body:
  i$_{\texttt{cur}}$ = $\phi$(base: i$_{\texttt{base}}$,
          $~$body: i$_{\texttt{body}}$)
  [...]
  i$_{\texttt{body}}$= i$_{\texttt{cur}}$+1
  [...]
  cond = ... // exit loop?
  br cond, out, body
out:
  x = i$_{\texttt{body}}$
\end{lstlisting}
\end{tabular}
\vspace{-1mm}
\caption{Original code} 
\end{subfigure}
\begin{subfigure}[b]{0.52\columnwidth}
\centering
\begin{tabular}{c}
\begin{lstlisting}[style=myVerbatimStyle]
base:
  i$_{\texttt{base}}$ = 0
body:
  i$_{\texttt{cur}}$ = $\phi$(base: i$_{\texttt{base}}$, body: i$_{\texttt{body}}$)
  i$_{\texttt{real}}$ = $\phi$(base: undef, body: i$_{\texttt{out}}$)
  i$_{\texttt{body}}$ = i$_{\texttt{cur}}$+1
  i$_{\texttt{out}}$= ct_select(taken, i$_{\texttt{body}}$, i$_{\texttt{real}}$)
  [...]
  cond = ... // exit loop?
  cfl_cond = ... // CFL override
  br cfl_cond, out, body
out:
  x = i$_{\texttt{out}}$
\end{lstlisting}
\end{tabular}
\vspace{-1mm}
\caption{After linearization} 
\end{subfigure}
\end{scriptsize}
\vspace{-0.75em}
\caption{Linearization with local variables outliving loops.\label{fig:loops}}
\vspace{-0.75em}
\end{figure}

After IR normalization, a loop is a single-entry, single-exit region: its exit block checks some condition $\textit{cond}$ for whether the program should leave the loop or take the back-edge to the loop \edit{body}. Note that break/continue statements \edit{are just} branches to the exit node and we linearize them as in \textsection\ref{ss:branches}. Before entering the loop we set $\textit{c\_idx}\mathbin{:=}0$, and modify the exit block in such a way that the program still makes the original $\textit{cond}$ computation, but uses instead the current $\textit{c\_idx}$ value to decide whether to leave the loop.

Say that we expect the  program to execute the loop no more than $k$ times (we address \edit{loop profiling} in \textsection\ref{ss:taint}). At every iteration our exiting decision procedure increments $\textit{c\_idx}$ by 1 and faces:

\begin{enumerate}
\item $\textit{taken}\mathbin{=}\textit{true}\wedge\textit{cond}\mathbin{=}\textit{false}$. The program is on a real path and wishes to take the back-edge \edit{to the body}: we allow it;
\item $\textit{taken}\mathbin{=}\textit{true}\wedge\textit{cond}\mathbin{=}\textit{true}$. The program is on a real path and wishes to exit the loop: if $\textit{c\_idx}=k$ we allow it, otherwise we enter dummy mode ($\textit{taken}\mathbin{:=}\textit{false}$) and the program will perform next $k\mathbin{-}\textit{c\_idx}$ dummy iterations for PC-security;
\item $\textit{taken}\mathbin{=}\textit{false}$. We make the program leave the loop when $\textit{c\_idx}=k$, and take the back-edge otherwise.
\end{enumerate}

Note that for (3) we do not use the value of $\textit{cond}$, as it can go rogue along decoy paths, but we still read it for the sake of linearization. Additionally, during (1) we validate the prediction of the oracle: whenever real program paths wish to iterate more than $k$ times, we adaptively update $k$ allowing the loop to continue, and use the $k'$ seen on loop exit as the new bound when the program reaches the loop again. The handling of this comparison is also linearized.

Nested loops or linearized branches in loop bodies pose no challenge: we incrementally update the taken predicate and restore it across regions as we did for nested branches in \textsection\ref{ss:branches} and Figure~\ref{fig:dummy}.

Let us resume the discussion of the code fragment. As variable $x$ outlives the first loop, we should prevent decoy paths from updating it for the sake of correctness. If the compiler places $x$ in memory, the IR will manipulate it using load and store instructions, and the dummy execution abstraction guarantees that only real paths can modify it. If instead it uses a virtual register $v$ for performance, we flank it with another register $v'$ conditionally assigned according to \taken, and replace all the uses of $v$ as operand in the remainder of the CFG with $v'$. Figure~\ref{fig:loops} shows this transformation with $\texttt{i}_{\texttt{body}}$ and $\texttt{i}_{\texttt{out}}$: decoy paths keep modifying $i_{\texttt{body}}$ for the sake of PC-security, but do not pollute the program state. Thanks to this design, we \edit{do not} demote $v$ to memory storage, which could harm performance especially for tight loops, \edit{nor} we constrain the optimizer. 


\subsubsection{Operand Sanitization}
\label{ss:var-latency}
As last step, we safeguard computations that could cause termination leaks from rogue values along decoy paths. In our design, this may happen only with divisions instructions receiving zero as divisor value. In \textsection\ref{se:background} we noted that x86 integer division is also subject to variable latencies from operand values. We address both issues \edit{via software emulation}, replacing {\tt *div} and {\tt *rem} LLVM instructions with subroutines that execute in constant-time, and for {\tt *div} are also insensitive to rogue values.

\subsubsection{Code Generation}
\label{ss:codegen}

Our CFL design poses no restrictions on code optimization as well as code generation operated \cready{in the backend. The optimizer can transform CFL-generated indirect memory references by means of optimizations such as common subexpression elimination and the code generator can lower such references using the most efficient patterns for the target architecture (including {\tt cmov} instructions on occasion). However, we need to prevent the code generation process from inadvertently adding branches in branchless IR-level code. Indeed, this is not uncommon~\cite{Molnar-ICISC05,binsecrel-SP20}: luckily, modern compilers offer explicit support to preserve our constant-time invariants. In more detail, we use LLVM backend options (e.g., {\tt -x86-cmov-converter=0} for branchless lowering on x86) to control this behavior.} As discussed later, we have also experimentally validated \proj{}-instrumented binaries preserve our security invariants by means of a dedicated verifier.

\subsection{Data Flow Linearization}
\label{se:dfl}

With data flow linearization (DFL), we devise a new abstraction for controlling the data access patterns influenced by secret data, so that arbitrarily different (secret) inputs will lead to the same observable program behavior for an attacker. As we discuss in our security evaluation of~\textsection\ref{se:security}, this design \edit{hardens against} side-channel attacks that prior solutions cannot handle \edit{and it does not suffer from leaks through data-flow invariants and memory safety violations as we saw for such solutions in~\textsection\ref{se:background}}. Furthermore, thanks to its combination with \cready{points-to} analysis, DFL is the first solution that \edit{does not place restrictions on pointer and object types, supporting for instance pointer-to-pointer casts that occur in real-world crypto code.}

\vspace{2pt}
{\centering
\begin{tcolorbox}[fontupper=\itshape\textsf, width=0.975\columnwidth,boxsep=0pt,top=6pt, bottom=6pt, left=10pt, right=10pt]
\small
\textbf{DFL}: For every program point that performs a memory load or store operation, DFL obliviously accesses all the locations that the original program can possibly reference for any initial input.
\end{tcolorbox}
\par}


To support this \scz{invariant} we conduct a context-sensitive, field-sensitive points-to analysis (described in \textsection\ref{ss:points-to}) to build DFL metadata for each use of a pointer expression in a sensitive load or store instruction. Such metadata describes the portions of the object(s) that the expression may reference each time the program evaluates it. {\color{black} We assume that only program-allocated memory can hold secret-dependent data (external library calls cannot leak from \textsection\ref{se:threatmodel})}.

For dynamic storage, that is stack- and heap-allocated objects, we instrument the involved allocation sites in the program to keep track at run time of the object instances currently stemming from an allocation site of interest to DFL (rightmost part of Figure~\ref{fig:architecture}). 

DFL uses indirection around incoming pointer values: it obliviously accesses all the candidate object portions identified by the \cready{points-to} analysis, and retrieves or modifies the memory value only within the object instance (if any) corresponding to the incoming pointer value. We apply DFL to every memory load or store made in a code region linearized by CFL (where the operation will see an incoming $\bot$ value when on a decoy path), and to memory operations that are outside input-dependent control flows but still secret-sensitive (e.g., array accesses with input-dependent index).

Unlike prior solutions, we do not need a shadow location for decoy paths (accessing it would leak the nature of such paths, \textsection\ref{se:background}), nor we let rogue pointers concur to memory accesses. Our design makes data accesses oblivious to secret dependencies and to the nature of control paths, \cready{and preserves} memory safety in the process}.




\begin{figure}[t!]
\begin{scriptsize}
\centering
\begin{tabular}{c}
\begin{lstlisting}[style=myVerbatimStyle]
typedef struct dfl_obj_list {
    struct dfl_obj_list* next;
    struct dfl_obj_list* prev;
    struct dfl_obj_list** head_ptr;  // for fast removal from list
    unsigned long magic;   // to distinguish DFL heap objects
    unsigned char data[];  // contents of program object
} dfl_obj_list_t;
\end{lstlisting}
\end{tabular}
\end{scriptsize}
\vspace{-1em}
\caption{In-band metadata for data flow linearization.\label{fig:dfl-metadata}}
\vspace{-1em}
\end{figure}

\subsubsection{Load and Store Wrappers}
\label{ss:dfl-wrappers}
For the linearization of the data flow of accessed locations, we use {\tt ct\_load} and {\tt ct\_store} primitives for DFL indirection and resort to different implementations optimized for the storage type and the size of the object instances to stride obliviously. As we \edit{discussed} when presenting the CFL stage, we accompany each use of a pointer expression in a load or store with DFL metadata specific to the program point.

DFL metadata capture at compilation time the {\em points-to information} for all the allocation sites of possibly referenced objects. The analysis comprises stack allocations, objects in global memory, and heap allocation operations. For each site, we use field-accurate information to limit striding only to portions of an object, which as a whole may hold thousands of bytes in real-world code. 

Depending on the scenario, the user can choose the granularity $\lambda$ at which memory accesses should become {\color{black} oblivious} to an adversary. One may only worry about cache attacks ($\lambda\mathbin{=}64$) if, say, only cross-core (cache) attacks are to be mitigated (e.g., with cloud vendors preventing core co-location across security domains by construction~\cite{phoronix-patches}). Or one may worry about arbitrary attacks if, say, core colocation across security domains is possible and attack vectors like MemJam ($\lambda\mathbin{=}4$) are at reach of the attacker.

%
Our wrapper implementations stride an object portion with a pointer expression incremented by $\lambda$ bytes every time and which may match the incoming $p$ input pointer from the program at most once. 
Depending on the object portion size, DFL picks between standard AVX instructions for striding, AVX2/AVX512 gather-scatter sequences to load many cache lines at once followed by custom selection masks, and a {\tt cmov}-based sequence that we devise to avoid the AVX setup latency for small objects (details in Appendix~\ref{apx:striding}).



The DFL load and store wrappers inspect all the allocation sites from the metadata. For global storage only a single object instance exists; for stack and heap objects the instances may change during the execution, and the wrappers inspect the run-time metadata that the transformed program maintains (using doubly linked lists and optimizations that we describe in the next sections).


For a load operation, DFL strides all the object instances that the program point may reference and conditionally selects the value from the object portion matching the desired address. For decoy paths, no match is found and {\tt ct\_load} returns a default value.

For a store operation, DFL breaks it into a load followed by a store. The rationale is to write to every plausible \scz{program point's target}, or the adversary may discover a secret-dependent write destination. For every object portion identified by DFL store metadata, we read the current value and replace it with the contents for the store only when the location matches its target, otherwise we write the current value back to memory. \scz{Decoy paths ``refresh'' the contents of each object; real paths do the same for all but the one they modify.}




\subsubsection{Object Lifetime}
\label{ss:obj-lifetime}
DFL metadata supplied at memory operations identify objects based on their allocation site and characteristics. While global storage is visible for the entire execution, stack and heap locations have a variable lifetime, and we need to maintain run-time metadata for their allocation sites.

We observe that real-world crypto code frequently allocates large structures on the stack and pointers seen at memory operations may reference more than one such structure. At the LLVM IR level, stack-allocated variables take the form of {\tt alloca} instructions that return a pointer to one or more elements of the desired type. The compiler automatically releases such storage on function return.

We interpose on {\tt alloca} to wrap the object with {\em in-band metadata} information depicted in Figure~\ref{fig:dfl-metadata}. Essentially, we prepend the originally allocated element with fields that optimize DFL operations and preserve stack alignment: the program element becomes the last field of a variable-sized {\tt dfl\_obj\_list\_t} structure. Then, we assign the virtual register meant to contain the $v$ pointer from {\tt alloca} with the address of $v.\textit{data}$ (32-byte offset on x64).

This transformation is simple when operating at the compiler IR level: unlike binary rewriting scenarios~\cite{Retrowrite-SP20}, the compiler is free to modify the stack layout while preserving program semantics, including well-behaved pointer arithmetics. Upon {\tt alloca} interposition, we make the program update the run-time allocation site information and a symmetric operation happens on function exit.

Heap variables see a similar treatment. We interpose on allocation operations to widen and prepend the desired object with in-band metadata, with the address of $v.\textit{data}$ returned to the program instead of the allocation base $v$. The $v.\textit{magic}$ field is pivotal for handling {\tt free()} operations efficiently: when interposing on them, we may witness a {\tt dfl\_obj\_list\_t} structure or a ``standard'' object from other program parts. We needed an efficient means to distinguish the two cases, as {\tt free()} operations take the allocation base as input: for DFL objects we have to subtract $32$ from the input pointer argument. We leverage the \edit{fact} that allocators like the standard libc allocator {\tt ptmalloc} prepend objects with at least one \edit{pointer}-sized field. Hence accessing a heap pointer $p$ as $p-8$ is valid: for DFL objects it would be the address of the $\textit{magic}$ field and we check its peculiar value to identify them. 


\subsubsection{Optimizations} 
\label{sss:dfl-opts}
One advantage of performing DFL \cready{at compiler IR level} is that we can further optimize both the data layout to ease metadata retrieval and the insertion of our DFL wrappers.

We identify functions that do not take part in recursive patterns and promote to global variables their stack allocations that sensitive accesses may reference. The promotion is sound as such a function can see only one active stack frame instance at a time. The promotion saves DFL the overhead of run-time bookkeeping, with faster metadata retrieval for memory operations as we discuss next. To identify functions apt for promotion, we analyze the call graph of the program (made only of direct calls after IR normalization) to identify strongly connected components from recursion patterns~\cite{scc-toplas97} and exclude functions taking part in them.

We also partially inline DFL handlers, as object allocation sites are statically known from \cready{points-to} analysis. For global storage, we also hard-code the involved address and striding information. For instance, a load operation from address $\textit{ptr}$ becomes in pseudo-code:


\vspace{-1.5pt}
{\centering
\begin{lstlisting}[language=C, style=inlineCode]
  res = 0
  res |= dfl_glob_load(ptr, glob1, stride_offset_g1, stride_size_g1)
  res |= dfl_glob_load(ptr, glob2, stride_offset_g2, stride_size_g2)
  res |= dfl_load(ptr, objs_as1, stride_offset_as1, stride_size_as2)
  res |= dfl_load(ptr, objs_as2, stride_offset_as2, stride_size_as2)
  res |= dfl_load(ptr, objs_as3, stride_offset_as2, stride_size_as2).
\end{lstlisting}
\par}

This is because the oracle determined that $\textit{ptr}$ may reference (portions of) global storage $\textit{glob1}$, $\textit{glob2}$ or objects from allocation sites $\textit{as1}$, $\textit{as2}$, $\textit{as3}$, where $\textit{objs\_as}_{i}$ is the pointer to the data structure (a doubly linked list of objects, as with AS2 in Figure~\ref{fig:architecture}) maintained at run time for the allocation site (\textsection\ref{ss:obj-lifetime}). With the OR operations we perform value selection, as each {\tt dfl\_} helper returns 0 unless the intended location $\textit{ptr}$ is met during striding. In other words, instead of maintaining points-to sets for memory operations as data, we inline their contents for performance (saving on retrieval time) and leave the LLVM optimizer free to perform further inlining of {\tt dfl\_} helpers code. The treatment of store operations is analogous.

\edit{Finally, we devise an effective (\textsection\ref{se:case-study}) striding optimization for loops. We encountered several loops where the induction variable flows in a pointer expression used to access an object, and from an analysis of its value (based on LLVM's {\em scalar evolution}) we could determine an invariant: the loop would be touching all the portions that require DFL striding and a distinct portion at each iteration. In other words, the code is ``naturally'' striding the object: we can avoid adding DFL striding and thus save on $n(n\mathbin{-}1)$ unnecessary accesses.}

%

\subsection{Support Analyses}


The \edit{compatibility of \proj with real-world code} stems also from two ``oracles'' as we tailor \edit{robust implementations of} mainstream program analysis techniques to our context: an {\em information flow tracking} technique to identify program portions affected by a secret and a {\em \cready{points-to} analysis} that we enhance with context sensitivity to obtain points-to sets as accurate as possible. 


\label{se:support}
\subsubsection{Identifying Sensitive Program Portions}
\label{ss:taint}
Control and data flow linearization need to be applied only to regions affected by secret data, \edit{as protecting non-leaky code only hampers performance.}

We assume the user has at their disposal a profiling suite to exercise the alternative control and data flow paths of the crypto functionality they seek to protect. Developers can resort to \edit{existing test suites for libraries}, actual workloads, or program testing tools (e.g. generic~\cite{aflpp-WOOT20} or specialized~\cite{ctfuzz} fuzzers) to build one. 

\cready{We then use DataFlowSanitizer (DFSan), a dynamic information flow tracking solution for LLVM, to profile the normalized IR of \textsection\ref{ss:ir-norm} over the profiling suite. DFSan comes with taint propagation rules for virtual registers and program memory and with APIs to define taint source and sink points. We write taint source configurations to automatically taint data that a program reads via I/O functions (e.g., a key file) and use as sink points conditionals, memory load/store operations, and {\tt div}/{\tt rem} instructions in the normalized IR. In the DFSan-transformed IR we then encode rules in the spirit of FlowTracker~\cite{FlowTracker-CC16} to handle implicit flows among virtual registers, leaving those possibly taking place through memory to complementary tools like FTI~\cite{GREYONE-SEC20}.}


We aggregate DFSan outputs to build a set of branches and memory accesses that are secret-dependent, feeding it to CFL and DFL. \edit{As we mentioned in \textsection\ref{se:cfl}, CFL will then push the hardening process to nested flows, linearizing their control and data flows.} During the execution of the profiling suite we also profile loop trip counts that we later use as initial predictions for CFL (\textsection\ref{ss:loops}).

\subsubsection{\cready{Points-to} Analysis}
\label{ss:points-to}
Points-to analyses~\cite{pointer-analysis-tutorial} determine the potential targets of pointers in a program. Nowadays they are available off-the-shelf in many compilation systems, with inclusion-based approaches in Andersen style~\cite{Andersen94} typically giving the most accurate results. In \proj, we extend the Andersen-style \cready{analysis} of the popular SVF library for LLVM~\cite{SVF-CC16}. For each pointer usage in the program, we use this analysis to build the {\em points-to set} of objects that it may reference at run time. Typically, \cready{points-to analyses collapse object instances from a dynamic allocation site} into an abstract single object\cready{. Hence,} points-to sets contain information on object allocation sites and static storage locations.

\cready{Points-to analyses are sound. However, they} may overapproximate sets by including objects that the program would never access at run time. In a lively area of research, many solutions feature inclusion-based analyses as the approach is more accurate than the alternative, faster unification-based one~\cite{Steensgaard-POPL96}.
%
Inclusion-based analyses could give even more accurate results if they were to scale to {\em context sensitivity}, i.e., they do not distinguish the uses of pointer expressions (and thus potentially involved objects) from different execution contexts. The context is typically intended as call-site sensitivity, while for object-oriented managed languages other definitions exist~\cite{Milanova-ISSTA02, Smaragdakis-POPL11}. To optimize the performance of DFL, we need as accurate points-to sets as possible, so in \proj we try to restore context sensitivity in an effective manner for a sufficiently large codebase such as the one of a real-world crypto library.



\vspace{-2pt}
\paragraph{Aggressive Cloning}
We use function cloning to turn a context-insensitive analysis in a context-sensitive one. A calling context~\cite{Delia-PLDI11} can be modeled as an acyclic path on the call graph and one can create a function clone for every distinct calling context encountered during a graph walk. This approach can immediately spin out of control, as the number of acyclic paths is often intractable~\cite{Whaley-PLDI04,hcct-spe}. 

Our scenario however is special, as we may clone only the functions identified as secret-dependent by the other oracle, along with their callees, recursively. We thus explore {\em aggressive cloning} along the maximal subtrees of the call graph having a sensitive function as root. The rationale is that we need maximum precision along the program regions that are secret-dependent, while we can settle for context-insensitive results for the remainder of the program, which normally dominates the codebase size.

Aggressive cloning turns out to be a key performance enabler, making DFL practical and saving on important overheads. As we discuss in \textsection\ref{se:case-study}, for wolfSSL we obtain points-to sets that are \textasciitilde6x smaller than the default ones of SVF and very close to the run-time optimum. The price that we trade for such performance is an increase in code size: this choice is common in much compiler~\pagebreak[4] research, both in static~\cite{Muth-SAS2000} and dynamic~\cite{Boisvert-CC10} compilation settings, for lucrative optimizations such as value and type-based specialization.


\vspace{-2pt}
\paragraph{Refined Field Sensitivity} 
A field-sensitive analysis can distinguish which portions of an object a pointer may reference. Real-world crypto code uses many-field, nested data structures of hundreds or thousands of bytes, and a load/store operation in the program typically references only a limited portion from them. Field-accurate information can make DFL striding cheaper: this factor motivated our practical enhancements to the field-sensitive part of SVF.

The reference implementation fails to recover field-precise information for about nine-tenths of the wolfSSL accesses that undergo DFL, especially when pointer arithmetics and optimizations are involved. We delay the moment when SVF falls back to field-insensitive abstract objects and try to reverse-engineer the structure of the addressing so to fit it into static type declarations of portions of the whole object. Our techniques are inspired by {\em duck typing} from compiler research; we cover them in Appendix~\ref{apx:field-heuristics}. Thanks to these refinements, we could recover field-sensitive information for pointers for {\color{black} 90\%} of the sensitive accesses in our case study.

\vspace{-2pt}
\paragraph{Indirect Calls} Points-to analysis also reveals possible targets for indirect calls~\cite{SVF-CC16}. We use this information during IR normalization \edit{\scz{when promoting them to if-series of guarded direct calls (\textsection\ref{ss:ir-norm})}, so to remove leaks from variable targets. We refine the candidates found by SVF at call sites by matching function prototype information and eliminating unfeasible targets. Indirect call target information is also necessary for the aggressive cloning strategy.}



\subsection{Discussion}
\label{ss:design-discussions}

\proj implements a compiler-based solution for eliminating microarchitectural side channels while coping with the needs of real-word code. We chose LLVM for its popularity and the availability of mature information-flow and points-to analyses. Nonetheless, our transformations are general and could be applied to other compilation toolchains. Similarly, we focus on x86/x64 architectures, but multiplexing conditional-assignment and SIMD striding instructions \scz{exist} for others as well (e.g. ARM SVE~\cite{ARMSVE-MICRO17} , RISC-V ``V''~\cite{RISCV-V}).

Moreover, operating at the compiler IR level allows us to efficiently add a level of indirection, with \taken unleashing the optimizer and with DFL making memory accesses oblivious to incoming pointers. In addition, aggressive function cloning allows us to transform the codebase and unveil a significantly more accurate number of objects to stride. The IR also retains type information that we can leverage to support field sensitivity and refine striding.

The just-in-time strategy to linearize secret-dependent unbounded control flows (loops) allows us to dodge intractability with high bounds and code bloat with tractable instances~\cite{Soares-CGO21}. For points-to set identification and indirect call promotion, our analyses yield very accurate results (i.e., closely matching the run-time accesses) on the programs we consider. We leave the exploration of a just-in-time flavor for them to future work, which may be helpful in non-cryptographic applications.


The main shortcoming of operating at the IR level is the inability to handle inline assembly sequences found in some crypto libraries. \cready{While snippets that break constant-time invariants are uncommon, they still need special handling, for instance with annotations or lifting. Verification-oriented lifting~\cite{bardin-ase19}, in particular, seems a promising avenue as it can provide formally verified C equivalent representations that we could use during IR normalization.}


As the programs we study do not exercise them, for space limitations we omit the treatment of recursion and multithreading. Appendix~\ref{apx:constructs} details the required implementation extensions.

\section{Security Analysis}
\label{se:security}

This section presents a security analysis of our transformations. We start by arguing that instrumented programs are semantically correct and induce secret-oblivious code and data access traces. We then discuss how our design emerges unscathed by traditional passive and active attacks and examine the residual attack surface.


\vspace{-2pt}
\paragraph{Correctness and Obliviousness}

Correctness follows directly from our design, as all our transformations are semantics-preserving. In short, for control flows, real paths perform all and only the computations the original program would make. For data flows, values from decoy paths cannot flow into real paths and correctness properties such as memory safety are preserved. Appendix~\ref{apx:correctness} provides informal proofs for these claims.

We now discuss how our linearization design yields \textit{oblivious} code and data access traces. For code accesses, PC-security follows by CFL construction, as we removed conditional branches, loops see a fixed number of iterations (we discuss wrong trip count predictions later), and IR normalization handles abort-like sequences. For data accesses, we wrapped load and store operations with DFL machinery that strides portions of every abstract (i.e., by allocation site) object that the operation may access, \textit{independently of the incoming pointer value}. For dynamic storage, for any two secrets, the program will see identical object collections to maintain at run-time: the composition of the lists can vary during the execution, but identically so for both secrets. Finally, for virtual registers that are spilled to memory by the backend, the CPU reads and writes them with the same instructions regardless of the current \taken predicate value, so those accesses are also oblivious.

\vspace{-2pt}
\paragraph{Security Properties} We build on the obliviousness claims above to show that both passive attacks (attackers only monitoring microarchitectural events) and active attacks (attackers also arbitrarily tampering with the microarchitectural state) are unsuccessful.

No instruction latency variance from secret-dependent operand values is possible, since we replace and sanitize instructions such as division (\textsection\ref{ss:var-latency}).
Memory accesses may have variable latencies, but, thanks to the DFL indirection, those will only depend on non-secret code/data and external factors. Moreover, DFL wrappers do not leak secrets and do not introduce decoy paths side channels in terms of decoy data flows or exceptions. In \textsection\ref{ss:dfl-wrappers}, we explained how load and store helpers stride objects using safe~\cite{Coppens-SP09, RACCOON-SEC15} {\tt cmov} or SIMD instructions. As for decoy paths, \taken can conditionally assign an incoming pointer with $\bot$: the adversary would need access to CPU registers or memory contents to leak the nature of a path (outside the threat model). Finally, helpers are memory-safe as points-to analysis is sound and we track object lifetimes.

Finally, an active attacker may perturb the execution to attempt Flush+Reload, Prime+Probe, and other microarchitectural attacks to observe cache line-sized or even word-sized victim accesses. With vulnerable code, they could alter for instance the access timing for a specific portion of memory, and observe timing differences to detect matching victim accesses. However, thanks to the obliviousness property of our approach, leaking victim accesses will have no value for the attacker, because we access all the possible secret-dependent code/data locations every time.

\vspace{-0.29em}
\paragraph{Residual Attack Surface}
We now discuss the residual attack surface for \proj. Design considerations aside, the correctness and obliviousness of the final instrumented binary are also subject to the correctness of our \proj implementation. Any implementation bug may introduce an attack surface. To mitigate this concern, we have validated our correctness claims experimentally by running extensive benchmarks and test suites for the programs we considered in our evaluation. We have also validated our obliviousness claims experimentally by means of a verifier, as detailed later. Overall, our implementation has a relatively small trusted computing base (TCB) of around 11~KLOC (631 LOC for our profiler, 955 LOC for CFL, 2561 for DFL, and 7259 LOC for normalization and optimization passes), which provides confidence it is possible to attain correctness and obliviousness in practice.

\proj's residual attack surface is also subject to the correctness of the required oracle information. The static points-to analysis we build on~\cite{SVF-CC16} is sound by design and our refinements preserve this property---barring again implementation bugs. Our information-flow tracking profiler, on the other hand, relies on the completeness of the original profiling suite to eliminate \scz{any} attack surface. While this is a fundamental limitation of dynamic analysis, we found straightforward to obtain the required coverage for a target secret-dependent computation, especially in cryptographic software. Implementation bugs or limitations such as implicit flows (\textsection\ref{ss:taint}) also apply here. A way to produce a more sound-by-design oracle is to adopt static information-flow tracking, but this also introduces overtainting and hence higher overheads~\cite{RACCOON-SEC15}.


An incomplete suite might also underestimate a secret-dependent loop bound. Thanks to just-in-time linearization correctness is not affected, but every time the \scz{trip count} is mispredicted (i.e., real-path loop execution yields a higher count than \scz{the oracle}), the adversary may observe a one-off perturbation (given that the instrumentation quickly adapts the padding). This is insufficient to leak any kind of high-entropy secret, but one can always envision pathological cases. Similar considerations can be applied to recursive functions.

In principle, other than statically unbound secret-dependent control flows, one can also envision statically unbound secret-dependent data flows such as a secret-dependent heap-allocated object size. We have not encountered such cases in practice, but they can also be handled using just-in-time (data-flow) linearization---i.e., padding to the maximum allocation size encountered thus far during profiling/production runs with similar characteristics.

\cready{Part of} the residual attack surface are all the code/data accesses fundamentally incompatible with linearization and constant-time programming in general. For instance, on the CFL front, one cannot linearize \cready{imbalanced if-else constructs that invoke system calls, or more generally} secret-dependent code paths executing arbitrary library/system calls. Their execution must remain conditional. A way to reduce the attack surface is to allow linearization of idempotent library/system calls or even to include some external library/system code in the instrumentation. On the DFL front, one cannot similarly linearize secret-dependent data accesses with external side effects, for instance those to a volatile data structure backed by a memory mapped I/O region (e.g., a user-level ION region~\cite{van_der_veen_drammer_2016}). Again, we have not encountered any of these pathological cases in practice.

\cready{
Similarly, \proj shares the general limitations of constant-time programming on the compiler and microarchitectural optimization front. Specifically, without specific provisions, a compiler backend may operate optimizations that inadvertently break constant-time invariants at the source (classic constant-time programming) or IR (automated solutions like \proj) level. Analogously, advanced microarchitectural optimizations may inadvertently re-introduce leaky patterns that break constant-time semantics. Some (e.g., hardware store elimination~\cite{hse}) may originate new instructions with secret-dependent latencies and require additional wrappers (and overhead). Others (e.g., speculative execution~\cite{kocher-sp19}) are more fundamental and require orthogonal mitigations.
}

\vspace{-2mm}
\section{Performance Evaluation}
\label{se:performance}
\vspace{-1pt}
This section evaluates \proj with classic benchmarks from prior work to answer the following questions: 
\begin{enumerate}[parsep=0pt,itemsep=0pt,topsep=1pt]
\item What is the impact of our techniques on compilation time?
\item How is binary size affected by linearization?
\item What are the run-time overheads induced by CFL and DFL?
\end{enumerate}
\vspace{1pt}

\vspace{-1mm}
\paragraph{Methodology}
We implemented \proj on top of LLVM 9.0 and SVF 1.9 and tested it on a machine with an Intel i7-7800X CPU (Skylake X) and 16 GB of RAM running Ubuntu 18.04. We discuss two striding configurations to conceal memory access patterns with DFL: word size ($\lambda\mathbin{=}4$), reflecting (core colocation) scenarios where recent intra cache level attacks like MemJam~\cite{MemJam} are possible, and cache line size ($\lambda\mathbin{=}64$), reflecting the common (cache attack) threat model of real-wold constant-time crypto implementations and also \proj's default configuration. We use AVX512 instructions to stride over large objects. Complete experimental results when using AVX2 and the $\lambda\mathbin{=}1$ configuration (presently out of reach for attackers) are further detailed in Appendix~\ref{apx:results}.
\cready{We study:
\begin{itemize}[topsep=1pt]
\item 23 realistic crypto modules manually extracted by the authors of SC-Eliminator~\cite{Wu-ISSTA18} from a 19-KLOC codebase (SCE suite), used also in the evaluation of Soares et al.~\cite{Soares-CGO21};
\item 6 microbenchmarks used in the evaluation of Raccoon~\cite{RACCOON-SEC15} (Raccoon suite)---all we could recover from the source code of prior efforts~\cite{GHOSTRIDER-ASPLOS15, MTO-CSF13}---using the same input sizes;
\item 8 targets used in constant-time verification works: 5 modules of the pycrypto suite analyzed in~\cite{data-usenix18} and 3 leaky functions of BearSSL and OpenSSL studied in Binsec/Rel~\cite{binsecrel-SP20}.
\end{itemize}
}
For profiling, we divide an input space of 32K elements in 128 equal partitions and pick a random instance from each, producing a profiling input set of 256 inputs. We build both the baseline and the instrumented version of each program at ({\tt -O3}). Table~\ref{tab:bench} presents our full experimental datasets with the SCE suite (first five blocks) \cready{and the Raccoon, pycrypto, and Binsec/Rel suites (one block each)}.

\vspace{-1mm}
\paragraph{Validation}
We validated the implementation for PC-security and memory access obliviousness with two verifiers. For code accesses, we use hardware counters for their total number and a cycle-accurate software simulation in GEM5. For data accesses, we use {\tt cachegrind} \cready{for cache line accesses} and \cready{write a DBI}~\cite{SoK-DBI} tool to track what locations an instruction accesses, including predicated {\tt cmov} ones visible at the microarchitectural level.
We repeatedly tested the instrumented programs in our datasets with~\pagebreak[4]  random variations of the profiling input set and random samples of the remaining inputs. We found no visible variations.

\paragraph{Compilation Time}
To measure \proj{}-induced compilation time, we applied our instrumentation to all the programs in our datasets and report statistics in Table~\ref{tab:bench}. The first four data columns report the sensitive program points identified with taint-based profiling over the randomly generated profiling input set. For the SCE programs, we protect the key scheduling and encryption stages.
\cready{For brevity, we report figures after cloning and after secret-dependent pushing to nested flows (\textsection\ref{se:cfl}): the former affected {\tt des3} and {\tt loki91}, while the latter affected {\tt applied-crypto/des}, {\tt dijkstra}, {\tt rsort}, and {\tt tls-rempad-luk13}.}
Interestingly, for {\tt 3way}, the LLVM optimizer already transformed out a secret-sensitive branch that would be visible at the source level, while no leaky data flows are present in it (consistently with~\cite{Wu-ISSTA18}).

Across all \cready{37} programs, the average dynamic analysis time for taint tracking and loop profiling was 4s, with a peak of 31.6s on {\tt libgcrypt/twofish} (\textasciitilde1 C KLOC). For static analysis (i.e., points-to), CFL/DFL transformations, and binary generation, the end-to-end average time per benchmark was 1.4s, with a peak of 23s on {\tt botan/twofish} (567 C++ LOC). Our results confirm \proj's instrumentation yields realistic compilation times.


\newcommand{\kt}[1]{\multicolumn{1}{@{\hskip3pt}l|@{\hskip3pt}}{#1}}

\begin{table}[t!]
\begin{scriptsize}
\centering
\caption{Benchmark characteristics and overheads.\label{tab:bench}}
\vspace{-1.6em}
\adjustbox{max width=0.925\columnwidth}{
\begin{tabular}{|@{\hskip2pt}c@{\hskip2pt}|
l@{\hskip2pt}|%
@{\hskip0pt}l@{\hskip3pt}|
@{\hskip3pt}l@{\hskip3pt}|%
@{\hskip3pt}l@{\hskip3pt}|%
@{\hskip3pt}l@{\hskip3pt}|%
@{\hskip0pt}r@{\hskip1pt}|%
@{\hskip1pt}r@{\hskip1pt}|%
@{\hskip0pt}r@{\hskip1pt}|%
@{\hskip1pt}r@{\hskip1pt}|} 
\cline{3-10}
\multicolumn{2}{c|}{} & \multicolumn{4}{c|}{\cready{IR constructs (sensitive/total)}} & \multicolumn{2}{c|}{performance} &\multicolumn{2}{c|}{binary size} \\
\cline{2-10}
\multicolumn{1}{c|}{} & program & \kt{branches}  & loops & reads & writes & {\tiny $\,\lambda\mathbin{=}4$} & {\tiny $\lambda\mathbin{=}64$} & {\tiny $\,\lambda\mathbin{=}4$} & {\tiny $\lambda\mathbin{=}64$} \\
\hline
\parbox[t]{0mm}{\multirow{8}{*}{\rotatebox[origin=t]{90}{\textsc{chronos}}}}
& aes & \kt{0/1} & 0/1 & 224/235 & 0/68 & 1.13x & 1.08x & 1.16x & 1.16x \\
& des &  \kt{0/1} & 0/1 & 318/362 & 0/36 & 1.19x & 1.14x & 1.37x & 1.73x \\
& des3 &  \kt{0/3} & 0/3 & 861/1005 & 0/89 & 1.49x & 1.36x & 1.92x & 2.84x \\
& anubis &  \kt{0/1} & 0/1 & 776/1240 & 0/87 & 1.29x & 1.12x & 1.27x & 1.27x \\
& cast5 &  \kt{0/1} & 0/1 & 333/372 & 0/36 & 1.13x & 1.06x & 1.16x & 1.16x \\
& cast6 &  \kt{-} & - & 192/204 & 0/4 & 1.13x & 1.08x & 1.01x & 1.01x \\
& fcrypt &  \kt{-} & - & 64/74 & 0/18 & 1.04x & 1.03x & 1.01x & 1.01x \\
& khazad &  \kt{-} & - & 136/141 & 0/1 & 1.13x & 1.09x & 1.15x & 1.15x \\
\hline
\parbox[t]{0mm}{\multirow{2}{*}{\rotatebox[origin=t]{90}{\textsc{s-cp}}}}
& aes\_core &  \kt{-} & - & 160/192 & 0/16 & 1.12x & 1.06x & 1.22x & 1.22x \\
& cast-ssl &  \kt{0/1} & 0/1 & 333/355 & 0/54 & 1.23x & 1.10x & 1.24x & 1.24x \\
\hline
\parbox[t]{0mm}{\multirow{6}{*}{\rotatebox[origin=t]{90}{\textsc{botan}}}}
& aes &  \kt{0/12} & 0/6 & 452/525 & 0/153 & 1.05x & 1.03x & 1.36x & 1.72x \\
& cast128 &  \kt{0/2} & 0/2 & 333/374 & 0/52 & 1.02x & 1.01x & 1.16x & 1.16x \\
& des &  \kt{0/1} & 0/1 & 136/185 & 0/24 & 1.01x & 1.01x & 1.16x & 1.16x \\
& kasumi &  \kt{0/7} & 0/7 & 96/174 & 0/18 & 1.01x & 1.01x & 1.29x & 1.57x \\
& seed &  \kt{0/6} & 0/6 & 320/360 & 0/41 & 1.02x & 1.01x & 1.18x & 1.18x \\
& twofish &  \kt{1/8} & 0/6 & 2402/2450 & 4/1084 & 1.14x & 1.12x & 1.45x & 2.42x \\
\hline
\parbox[t]{0mm}{\multirow{3}{*}{\rotatebox[origin=t]{90}{\textsc{app-cr}}}}
& 3way &  \kt{0/4} & 0/4 & 0/8 & 0/14 & 1.00x & 1.00x & 1.00x & 1.00x \\
& des &  \kt{2/10} & 0/6 & 134/182 & 2/17 & 1.24x & 1.09x & 1.23x & 1.45x \\
& loki91 &  \kt{16/76} & 24/28 & 16/24 & 0/6 & 1.51x & 1.43x & 1.02x & 1.02x \\
\hline
\parbox[t]{0mm}{\multirow{4}{*}{\rotatebox[origin=t]{90}{\textsc{libgcrypt}}}}
& camellia &  \kt{-} & - & 32/48 & 0/48 & 1.02x & 1.01x & 1.01x & 1.01x \\
& des &  \kt{0/2} & 0/2 & 144/195 & 0/12 & 1.06x & 1.06x & 1.29x & 1.85x \\
& seed &  \kt{0/4} & 0/1 & 200/265 & 0/18 & 1.18x & 1.10x & 1.22x & 1.22x \\
& twofish &  \kt{-} & - & 2574/2662 & 0/1080 & 1.97x & 1.92x & 1.43x & 2.24x \\
\hline
\parbox[t]{0mm}{\multirow{6}{*}{\rotatebox[origin=t]{90}{\textsc{raccoon}}}}
& binsearch &  \kt{1/4} & 1/2 & 1/3 & 0/2 & 1.33x & 1.18x & 1.01x & 1.01x \\
& dijkstra &  \kt{3/15} & 0/5 & 5/10 & 3/7 & 3.45x & 1.51x & 1.01x & 1.01x \\
& findmax &  \kt{0/2} & 0/2 & 0/1 & 0/1 & 1.00x & 1.00x & 1.00x & 1.00x \\
& histogram &  \kt{0/2} & 0/2 & 1/2 & 1/1 & 2.66x & 1.68x & 1.01x & 1.01x \\
& matmul &  \kt{0/5} & 0/5 & 0/2 & 0/2 & 1.00x & 1.00x & 1.00x & 1.00x \\
& rsort &  \kt{0/9} & 4/6 & 6/8 & 4/4 & 1.87x & 1.84x & 1.30x & 1.30x \\
\hline
\parbox[t]{0mm}{\multirow{5}{*}{\rotatebox[origin=t]{90}{\cready{\textsc{pycrypto}}}}}
& aes      &  \kt{0/11} & 0/5 & 96/223  & 0/59 & 1.13x & 1.06x & 1.19x & 1.19x \\
& arc4     &  \kt{0/3} & 0/3 & 3/30   & 2/10   & 1.07x & 1.03x & 1.01x & 1.01x \\
& blowfish &  \kt{0/16} & 0/12 & 24/77  & 0/39 & 5.07x & 3.17x & 1.01x & 1.01x \\
& cast     &  \kt{0/29} & 0/2 & 284/321 & 0/57 & 1.09x & 1.04x & 1.37x & 1.37x \\
& des3     &  \kt{0/5} & 0/1 & 32/40  & 0/7    & 1.06x & 1.04x & 1.01x & 1.01x \\
\hline
\parbox[t]{0mm}{\multirow{3}{*}{\rotatebox[origin=t]{90}{\cready{\textsc{B/Rel}}}}}
& tls-rempad-luk13 &  \kt{4/17} & 1/1 & 6/14  & 4/17   & 1.01x & 1.01x & 1.02x & 1.02x \\
& aes\_big         &  \kt{0/45} & 0/8 & 32/141 & 0/40  & 1.01x & 1.01x & 1.29x & 1.29x \\
& des\_tab         &  \kt{0/50} & 0/28 & 8/164  & 0/97 & 1.04x & 1.02x & 1.29x & 1.29x \\
\hline
\hline
\multirow{5}{*}{\shortstack[c]{\textsc{avg}\\ \tiny\textsc{(geo)}}} & SCE suite &  \kt{-} & - & - & - & 1.16x & 1.11x & 1.22x & 1.35x \\
& Raccoon suite & \kt{-}  & - & - & - & 1.68x & 1.33x & 1.05x & 1.05x \\
& \cready{pycrypto suite} & \kt{-}  & - & - & - & 1.48x & 1.30x & 1.10x & 1.10x \\
& \cready{Binsec/Rel suite} & \kt{-}  & - & - & - & 1.02x & 1.01x & 1.19x & 1.19x \\
& \cready{all programs} & \kt{-}  & - & - & - & \cready{1.26x} & \cready{1.16x} & \cready{1.17x} & \cready{1.25x} \\
\hline
\end{tabular}
}
\end{scriptsize}
\vspace{-1.15em}
\end{table}

\vspace{-1mm}
\paragraph{Binary Size}
Next, we study how our instrumentation impacts the final binary size. Two design factors are at play: cloning for the sake of accurate points-to information and DFL metadata inlining to avoid run-time lookups for static storage. Compared to prior solutions, however, we save instructions by avoiding loop unrolling.

During code generation, we leave the choice of inlining AVX striding sequences to the compiler, suggesting it for single accesses and for small stride sizes with the {\tt cmov}-based method of Appendix~\ref{apx:striding}---we observed lower run-time overhead from such choice. When we use word-level striding ($\lambda\mathbin{=}4$), the binary size is typically smaller than for cache line-level striding ($\lambda\mathbin{=}64$), as the AVX helpers for fast cache line accesses feature more complex logics.

As shown in Table~\ref{tab:bench}, the average binary size increment on the SCE suite is around 1.35x in our default configuration ($\lambda\mathbin{=}64$) and 1.22x for $\lambda\mathbin{=}4$. For {\tt des3}, we observe 1.92-2.84x increases mainly due to cloning combined with inlining. Smaller increases can be noted for the two {\tt twofish} variants, due to DFL helpers inlined in the many sensitive read operations. The binary size increase for all the other programs is below 2x. The Raccoon programs see hardly noticeable differences with the exception of {\tt rsort}, for which we observe a 1.3x increase. \cready{We note similar peak values in the two other suites, with a 1.37x increase for {\tt cast} in pycrypto and 1.29x for {\tt aes\_big} and {\tt des\_tab} in Binsec/Rel.} Our results confirm \proj's instrumentation yields realistic binary sizes.

\paragraph{Run-time Performance}

Finally, we study \proj's run-time performance. To measure the slowdown induced by \proj on our benchmarks, we measured the time to run each instrumented program by means of CPU cycles with thread-accurate CPU hardware counters (akin~\cite{Wu-ISSTA18}). We repeated the experiments 1,000 times and report the mean normalized execution time compared against the baseline. Table~\ref{tab:bench} presents our results.

\proj's default configuration produces realistic overheads across all our benchmarks, for instance with a geomean overhead of 11\% on the SCE suite and 33\% on the Raccoon programs. These numbers only increase to 16\% and 68\% for word-level protection. Our SCE suite numbers are comparable to those of SC-Eliminator~\cite{Wu-ISSTA18} and Soares et al.~\cite{Soares-CGO21} (which we confirmed using the artifacts publicly released with both papers, \cready{Appendix}~\ref{apx:results}), despite \proj offering much stronger compatibility (i.e., real-world program support) and security (i.e., generic data-flow protection and no decoy path side channels) guarantees. On the Raccoon test suite, on the other hand, Raccoon reported two orders-of-magnitude slowdowns (up to 432x) on a number of benchmarks, while \proj's worst-case slowdown in its default configuration is only 1.84x, despite \proj again providing stronger compatibility and security guarantees (i.e., no decoy path side channels). Overall, \proj significantly outperforms state-of-the-art solutions in the performance/security dimension on their own datasets, while providing much better compatibility with real-world programs. \cready{For the two other suites, we observe modest overheads with the exception of {\tt blowfish}: its 3.17-5.07x slowdown originates in a hot tight loop making four secret-dependent accesses on four very large tables, a pathological case of leaky design for automatic repair.}

\section{Case Study}
\label{se:case-study}

The wolfSSL library is a portable SSL/TLS implementation written in C and compliant with the FIPS 140-2 criteria from the U.S. government. It makes for a compelling case study for several reasons.

From a technical perspective, it is representative of the common programming idioms in real-world programs and is a complex, stress test for any constant-time programming solution (which, in fact, none of the existing solutions can even partially support). As a by-product, it also allows us to showcase the benefits of our design.

The library supports Elliptic Curve (EC) cryptography, which is appealing as it allows smaller keys for equivalent guarantees of non-EC designs (e.g. RSA)~\cite{nsa-faq}. EC Digital Signature Algorithms (ECDSA) are among the most popular DSA schemes today, yet their implementations face pitfalls and vulnerabilities that threaten their security, as shown by recent attacks such as LadderLeak~\cite{LADDERLEAK-CCS20} (targeting the Montgomery ladder behind the EC scalar multiplication in ECDSA) and CopyCat~\cite{COPYCAT-SEC20} (targeting the vulnerable hand-crafted constant-time (CT) wolfSSL ECDSA code).

In this section, we harden with \proj the {\tt mulmod} modular multiplication procedure in ECDSA from the non-CT wolfSSL implementation. This procedure calculates a curve point $k\times G$, where $k$ is a crypto-secure nonce and $G$ is the EC base point. Leaks involving $k$ bits have historically been abused in the wild for, e.g., stealing Bitcoin wallets~\cite{bitcoin2013} and hacking consoles~\cite{console-ccc10}.

\vspace{-1mm}
\paragraph{Code Features and Analysis}
The region to protect comprises 84 functions from the maximal tree that {\tt mulmod} spans in the call graph. We generate a profiling set of 1024 random inputs with 256-bit key length and identify sensitive branches, loops, and memory accesses (Table~\ref{tab:wolf}). The analysis of loops is a good example of how unrolling is unpractical. We found an outer loop iterating over the key bits, then 1 inner loop at depth 1, 4 at depth 2, and 3 at depth 3 (all within the same outer loop). Every inner loop iterates up to 4 times, resulting in a nested structure---and potential unroll factor---of 61,440. And this calulation is entirely based on profiling information, the inner loops are actually unbounded from static analysis.

Similarly, cloning is crucial for the accuracy of DFL. We profiled the object sets accessed at each program point with our DBI tool (\textsection\ref{se:performance}). With cloning, on average, a protected access over-strides (i.e., striding bytes that the original program would not touch) by as little as 8\% of the intended storage. Without cloning, on the other hand, points-to sets are imprecise enough that DFL needs to make as many as 6.29x more accesses than strictly needed.

\newcolumntype{L}[1]{>{\raggedright\let\newline\\\arraybackslash\hspace{0pt}}m{#1}}
\begin{table}[t!]
\begin{footnotesize}
\centering
\caption{Characteristics and overheads for wolfSSL.\label{tab:wolf}}
\vspace{-1.5em}
\adjustbox{max width=0.84\columnwidth}{ 
\begin{tabular}{r}
\begin{tabular}{|l|L{1.5cm}|L{1.5cm}|L{1.5cm}|}
\cline{2-4}
\multicolumn{1}{c|}{} & baseline & w/o cloning & w/ cloning \\
\hline
functions & 84 & 84 & 864 \\
binary size (KB) & 39 & 135 (3.5x) & 638 (16.35x) \\
exec cycles (M) & 2.6 & 200 (77x) & 33 (12.7x) \\ 
accessed objs/point & 1 & 6.29 & 1.08 \\
\hline
\end{tabular}%
\vspace{0.25em}
\\
\begin{tabular}{|l|L{1.5cm}|L{1.5cm}|L{1.5cm}|}
\cline{2-4}
\multicolumn{1}{c|}{} &
tainted & 
nested flows \tiny{(w/o cloning)} &
w/ cloning \tiny{nested (tainted)} \\
\hline
branches & 13 & 39 & 1046 (118) \\
loops & 12 & 31 & 863 (139) \\
reads & 33 & 138 & 2898 (52) \\ 
writes & 1 & 91 & 1892 (2) \\ 
\hline
\end{tabular}%
\vspace{0.25em}
\\
\begin{tabular}{|l|L{1.5cm}|L{1.5cm}|L{1.5cm}|}
\cline{2-4}
\multicolumn{1}{c|}{} & time {\tiny (ms)} & cycles {\tiny (M)} & binary size {\tiny (KB)} \\
\hline
wolfSSL (W=4) & 0.35 & 1.6 & 39 \\
wolfSSL (W=1) & 0.57 & 2.6 & 39 \\
wolfSSL (const. time) & 0.7 & 2.9 & 47 \\
\proj (W=1) & 8 & 33 & 638 \\
\hline
\end{tabular}%
\end{tabular}
}
\end{footnotesize}
\vspace{-1em}
\end{table}

\vspace{-1mm}
\paragraph{Overheads}

Table~\ref{tab:wolf} presents our run-time performance overhead results, measured and reported in the same way as our earlier benchmark experiments. As shown in the table, the slowdown compared to the original non-CT baseline of wolfSSL (using the compilation parameter W=1) is 12.7x, which allows the \proj{}-instrumented version to complete a full run in 8~ms. The compilation parameter W allows the non-CT version to use different double-and-add interleavings over the key bits as part of its sliding window-based double-and-add approach to implement ECC multiplication. In brief, a higher W value trades run-time storage (growing exponentially with W) with steady-state throughput (increasing linearly with W), but also alters the code generated, due to snowball optimization from inlined constants. This choice turns out to be cost-effective in the non-CT world, but not for linearization.


For completeness, we also show results for the best configuration of the non-CT version (which we profiled to be W=4) and the hand-written CT version of wolfSSL. The non-CT code completes an ECC multiplication in 0.35 ms in its best-performing scenario, while the hand-written CT version completes in 0.7 ms. Our automatically hardened code completes in 8ms, that is within a 11.42x factor of the hand-written CT version, using 11.38x more CPU cycles, yet with strong security guarantees for both control and data flows from the articulate computation (i.e., 84 functions) involved.

In terms of binary size increase, with cloning we trade space usage for DFL performance. We obtain a 16.36x increase compared to the reference non-CT implementation, and 13.57x higher size than the CT version. The performance benefits from cloning are obvious (77x/12.7x=6.06x end-to-end speedup) and the size of the binary we produce is 638 KB, which, in absolute terms is acceptable, but amenable to further reduction. In particular, the nature of wolfSSL code is tortured from a cloning perspective: it comprises 36 arithmetic helper functions that we clone at multiple usage sites. We measured, however, that in several cases they are invoked in function instances (which now represents distinct calling contexts for the original program) that see the same points-to information. Hence, after cloning, one may attempt merging back functions from calling contexts that see the same points-to set, saving a relevant fraction of code boat without hampering DFL performance.

Other optimizations, such as our DFL loop optimization also yields important benefits, removing unnecessary striding in some loops---without it, the slowdown would more than double (27.1x). We conclude by reporting a few statistics on analysis and compilation time. The profiling stage took 10m34s, the points-to analysis 20s (\textasciitilde2s w/o cloning), and the end-to-end code transformation and compilation process 1m51s (31s for the non-CT reference).

Overall, our results confirm that \proj can effectively handle a real-world crypto library for the first time, with no annotations to aid compatibility and with realistic compilation times, binary sizes, and run-time overheads. \proj's end-to-end run-time overhead, in particular, is significantly (i.e., up to two orders of magnitude) lower than what prior comprehensive solutions like Raccoon~\cite{RACCOON-SEC15} have reported on much simpler benchmarks.

\vfill 
\section{Conclusion}
We have presented \proj, an automatic constant-time system to harden programs against microarchitectural side channels. Thanks to carefully designed compiler transformations and optimizations, we devised a radical design point---complete linearization of control and data flows---as an efficient and compatible solution that brings security by construction, and can handle for the very first time a production-ready crypto library component. 

\section*{Acknowledgements}

We thank our shepherd Qi Li and the anonymous reviewers for their valuable feedback. This work was supported by the European Union's Horizon 2020 research and innovation programme under grant agreements No. 786669 (ReAct) and 825377 (UNICORE), and by Intel Corporation through the Side Channel Vulnerability ISRA.



\clearpage
\bibliographystyle{ACM-Reference-Format}
\bibliography{biblio}


\begin{thebibliography}{87}


\ifx \showCODEN    \undefined \def \showCODEN     #1{\unskip}     \fi
\ifx \showDOI      \undefined \def \showDOI       #1{#1}\fi
\ifx \showISBNx    \undefined \def \showISBNx     #1{\unskip}     \fi
\ifx \showISBNxiii \undefined \def \showISBNxiii  #1{\unskip}     \fi
\ifx \showISSN     \undefined \def \showISSN      #1{\unskip}     \fi
\ifx \showLCCN     \undefined \def \showLCCN      #1{\unskip}     \fi
\ifx \shownote     \undefined \def \shownote      #1{#1}          \fi
\ifx \showarticletitle \undefined \def \showarticletitle #1{#1}   \fi
\ifx \showURL      \undefined \def \showURL       {\relax}        \fi
\providecommand\bibfield[2]{#2}
\providecommand\bibinfo[2]{#2}
\providecommand\natexlab[1]{#1}
\providecommand\showeprint[2][]{arXiv:#2}

\bibitem[\protect\citeauthoryear{??}{con}{2010}]%
        {console-ccc10}
 \bibinfo{year}{2010}\natexlab{}.
\newblock \showarticletitle{Console Hacking 2010}.
\newblock  (\bibinfo{date}{Dec.} \bibinfo{year}{2010}).
\newblock
\urldef\tempurl%
\url{https://fahrplan.events.ccc.de/congress/2010/Fahrplan/events/4087.en.html}
\showURL{%
\tempurl}


\bibitem[\protect\citeauthoryear{??}{bit}{2013}]%
        {bitcoin2013}
 \bibinfo{year}{2013}\natexlab{}.
\newblock \showarticletitle{Bitcoin - Android Security Vulnerability}.
\newblock  (\bibinfo{date}{Aug.} \bibinfo{year}{2013}).
\newblock
\urldef\tempurl%
\url{https://bitcoin.org/en/alert/2013-08-11-android}
\showURL{%
\tempurl}


\bibitem[\protect\citeauthoryear{??}{mov}{2015}]%
        {movfuscator}
 \bibinfo{year}{2015}\natexlab{}.
\newblock \showarticletitle{The M/o/Vfuscator}.
\newblock  (\bibinfo{date}{Oct.} \bibinfo{year}{2015}).
\newblock
\urldef\tempurl%
\url{https://github.com/xoreaxeaxeax/movfuscator}
\showURL{%
\tempurl}


\bibitem[\protect\citeauthoryear{??}{RIS}{2019}]%
        {RISCV-V}
 \bibinfo{year}{2019}\natexlab{}.
\newblock \showarticletitle{{RISC-V} {"V"} Vector Extension}.
\newblock  (\bibinfo{date}{Nov.} \bibinfo{year}{2019}).
\newblock
\urldef\tempurl%
\url{https://riscv.github.io/documents/riscv-v-spec/riscv-v-spec.pdf}
\showURL{%
\tempurl}


\bibitem[\protect\citeauthoryear{??}{pho}{2020}]%
        {phoronix-patches}
 \bibinfo{year}{2020}\natexlab{}.
\newblock \showarticletitle{Google Publishes Latest Linux Core Scheduling
  Patches So Only Trusted Tasks Share A Core}.
\newblock  (\bibinfo{date}{Nov.} \bibinfo{year}{2020}).
\newblock
\urldef\tempurl%
\url{https://www.phoronix.com/scan.php?page=news_item&px=Google-Core-Scheduling-v9#:~:text=Google%20engineer%20Joel%20Fernandes%20sent,against%20the%20possible%20security%20exploits}
\showURL{%
\tempurl}


\bibitem[\protect\citeauthoryear{Agat}{Agat}{2000}]%
        {Agat-POPL00}
\bibfield{author}{\bibinfo{person}{Johan Agat}.}
  \bibinfo{year}{2000}\natexlab{}.
\newblock \showarticletitle{Transforming out Timing Leaks}. In
  \bibinfo{booktitle}{\emph{Proceedings of the 27th ACM SIGPLAN-SIGACT
  Symposium on Principles of Programming Languages}} (Boston, MA, USA)
  \emph{(\bibinfo{series}{POPL '00})}. \bibinfo{publisher}{Association for
  Computing Machinery}, \bibinfo{address}{New York, NY, USA},
  \bibinfo{pages}{40--53}.
\newblock
\showISBNx{1581131259}
\urldef\tempurl%
\url{https://doi.org/10.1145/325694.325702}
\showDOI{\tempurl}


\bibitem[\protect\citeauthoryear{Aho, Lam, Sethi, and Ullman}{Aho
  et~al\mbox{.}}{2006}]%
        {ullman06}
\bibfield{author}{\bibinfo{person}{Alfred~V. Aho}, \bibinfo{person}{Monica~S.
  Lam}, \bibinfo{person}{Ravi Sethi}, {and} \bibinfo{person}{Jeffrey~D.
  Ullman}.} \bibinfo{year}{2006}\natexlab{}.
\newblock \bibinfo{booktitle}{\emph{Compilers: Principles, Techniques, and
  Tools (2nd Edition)}}.
\newblock \bibinfo{publisher}{Addison-Wesley Longman Publishing Co., Inc.},
  \bibinfo{address}{USA}.
\newblock
\showISBNx{0321486811}


\bibitem[\protect\citeauthoryear{{Aldaya}, {Brumley}, {ul Hassan}, {Pereida
  García}, and {Tuveri}}{{Aldaya} et~al\mbox{.}}{2019}]%
        {Aldaya-sp19}
\bibfield{author}{\bibinfo{person}{A.~C. {Aldaya}}, \bibinfo{person}{B.~B.
  {Brumley}}, \bibinfo{person}{S. {ul Hassan}}, \bibinfo{person}{C. {Pereida
  García}}, {and} \bibinfo{person}{N. {Tuveri}}.}
  \bibinfo{year}{2019}\natexlab{}.
\newblock \showarticletitle{Port Contention for Fun and Profit}. In
  \bibinfo{booktitle}{\emph{2019 IEEE Symposium on Security and Privacy (SP)}}.
  \bibinfo{pages}{870--887}.
\newblock
\urldef\tempurl%
\url{https://doi.org/10.1109/SP.2019.00066}
\showDOI{\tempurl}


\bibitem[\protect\citeauthoryear{Andersen}{Andersen}{1994}]%
        {Andersen94}
\bibfield{author}{\bibinfo{person}{Lars~Ole Andersen}.}
  \bibinfo{year}{1994}\natexlab{}.
\newblock \emph{\bibinfo{title}{Program Analysis and Specialization for the C
  Programming Language}}.
\newblock \bibinfo{thesistype}{Ph.D. Dissertation}.
\newblock


\bibitem[\protect\citeauthoryear{Andrysco, Kohlbrenner, Mowery, Jhala, Lerner,
  and Shacham}{Andrysco et~al\mbox{.}}{2015}]%
        {Andrysco-SP15}
\bibfield{author}{\bibinfo{person}{Marc Andrysco}, \bibinfo{person}{David
  Kohlbrenner}, \bibinfo{person}{Keaton Mowery}, \bibinfo{person}{Ranjit
  Jhala}, \bibinfo{person}{Sorin Lerner}, {and} \bibinfo{person}{Hovav
  Shacham}.} \bibinfo{year}{2015}\natexlab{}.
\newblock \showarticletitle{On Subnormal Floating Point and Abnormal Timing}.
  In \bibinfo{booktitle}{\emph{Proceedings of the 2015 IEEE Symposium on
  Security and Privacy}} \emph{(\bibinfo{series}{SP '15})}.
  \bibinfo{publisher}{IEEE Computer Society}, \bibinfo{address}{USA},
  \bibinfo{pages}{623--639}.
\newblock
\showISBNx{9781467369497}
\urldef\tempurl%
\url{https://doi.org/10.1109/SP.2015.44}
\showDOI{\tempurl}


\bibitem[\protect\citeauthoryear{Andrysco, N\"{o}tzli, Brown, Jhala, and
  Stefan}{Andrysco et~al\mbox{.}}{2018}]%
        {CFTP-CCS18}
\bibfield{author}{\bibinfo{person}{Marc Andrysco}, \bibinfo{person}{Andres
  N\"{o}tzli}, \bibinfo{person}{Fraser Brown}, \bibinfo{person}{Ranjit Jhala},
  {and} \bibinfo{person}{Deian Stefan}.} \bibinfo{year}{2018}\natexlab{}.
\newblock \showarticletitle{Towards Verified, Constant-Time Floating Point
  Operations}. In \bibinfo{booktitle}{\emph{Proceedings of the 2018 ACM SIGSAC
  Conference on Computer and Communications Security}} (Toronto, Canada)
  \emph{(\bibinfo{series}{CCS '18})}. \bibinfo{publisher}{Association for
  Computing Machinery}, \bibinfo{address}{New York, NY, USA},
  \bibinfo{pages}{1369--1382}.
\newblock
\showISBNx{9781450356930}
\urldef\tempurl%
\url{https://doi.org/10.1145/3243734.3243766}
\showDOI{\tempurl}


\bibitem[\protect\citeauthoryear{Aranha, Novaes, Takahashi, Tibouchi, and
  Yarom}{Aranha et~al\mbox{.}}{2020}]%
        {LADDERLEAK-CCS20}
\bibfield{author}{\bibinfo{person}{Diego~F. Aranha},
  \bibinfo{person}{Felipe~Rodrigues Novaes}, \bibinfo{person}{Akira Takahashi},
  \bibinfo{person}{Mehdi Tibouchi}, {and} \bibinfo{person}{Yuval Yarom}.}
  \bibinfo{year}{2020}\natexlab{}.
\newblock \showarticletitle{{LadderLeak}: Breaking {ECDSA} with Less than One
  Bit of Nonce Leakage}. In \bibinfo{booktitle}{\emph{Proceedings of the 2020
  ACM SIGSAC Conference on Computer and Communications Security}}
  \emph{(\bibinfo{series}{CCS '20})}. \bibinfo{publisher}{Association for
  Computing Machinery}, \bibinfo{address}{New York, NY, USA},
  \bibinfo{pages}{225--242}.
\newblock
\showISBNx{9781450370899}
\urldef\tempurl%
\url{https://doi.org/10.1145/3372297.3417268}
\showDOI{\tempurl}


\bibitem[\protect\citeauthoryear{{Barthe}, {Gr{\'e}goire}, and
  {Laporte}}{{Barthe} et~al\mbox{.}}{2018}]%
        {Barthe18}
\bibfield{author}{\bibinfo{person}{G. {Barthe}}, \bibinfo{person}{B.
  {Gr{\'e}goire}}, {and} \bibinfo{person}{V. {Laporte}}.}
  \bibinfo{year}{2018}\natexlab{}.
\newblock \showarticletitle{Secure Compilation of Side-Channel Countermeasures:
  The Case of Cryptographic ''Constant-Time''}. In
  \bibinfo{booktitle}{\emph{2018 IEEE 31st Computer Security Foundations
  Symposium (CSF)}}. \bibinfo{pages}{328--343}.
\newblock
\urldef\tempurl%
\url{https://doi.org/10.1109/CSF.2018.00031}
\showDOI{\tempurl}


\bibitem[\protect\citeauthoryear{Bosman, Razavi, Bos, and Giuffrida}{Bosman
  et~al\mbox{.}}{2016}]%
        {bosman2016dedup}
\bibfield{author}{\bibinfo{person}{Erik Bosman}, \bibinfo{person}{Kaveh
  Razavi}, \bibinfo{person}{Herbert Bos}, {and} \bibinfo{person}{Cristiano
  Giuffrida}.} \bibinfo{year}{2016}\natexlab{}.
\newblock \showarticletitle{Dedup Est Machina: Memory Deduplication as an
  Advanced Exploitation Vector}. In \bibinfo{booktitle}{\emph{2016 IEEE
  Symposium on Security and Privacy (SP)}}. \bibinfo{pages}{987--1004}.
\newblock
\urldef\tempurl%
\url{https://doi.org/10.1109/SP.2016.63}
\showDOI{\tempurl}


\bibitem[\protect\citeauthoryear{Canella, Genkin, Giner, Gruss, Lipp, Minkin,
  Moghimi, Piessens, Schwarz, Sunar, Van~Bulck, and Yarom}{Canella
  et~al\mbox{.}}{2019}]%
        {Canella-ccs19}
\bibfield{author}{\bibinfo{person}{Claudio Canella}, \bibinfo{person}{Daniel
  Genkin}, \bibinfo{person}{Lukas Giner}, \bibinfo{person}{Daniel Gruss},
  \bibinfo{person}{Moritz Lipp}, \bibinfo{person}{Marina Minkin},
  \bibinfo{person}{Daniel Moghimi}, \bibinfo{person}{Frank Piessens},
  \bibinfo{person}{Michael Schwarz}, \bibinfo{person}{Berk Sunar},
  \bibinfo{person}{Jo Van~Bulck}, {and} \bibinfo{person}{Yuval Yarom}.}
  \bibinfo{year}{2019}\natexlab{}.
\newblock \showarticletitle{Fallout: Leaking Data on Meltdown-Resistant CPUs}.
  In \bibinfo{booktitle}{\emph{Proceedings of the 2019 ACM SIGSAC Conference on
  Computer and Communications Security}} \emph{(\bibinfo{series}{CCS '19})}.
  \bibinfo{publisher}{Association for Computing Machinery},
  \bibinfo{address}{New York, NY, USA}, \bibinfo{pages}{769--784}.
\newblock
\showISBNx{9781450367479}
\urldef\tempurl%
\url{https://doi.org/10.1145/3319535.3363219}
\showDOI{\tempurl}


\bibitem[\protect\citeauthoryear{Cauligi, Disselkoen, Gleissenthall, Tullsen,
  Stefan, Rezk, and Barthe}{Cauligi et~al\mbox{.}}{2020}]%
        {Cauligi-PLDI20}
\bibfield{author}{\bibinfo{person}{Sunjay Cauligi}, \bibinfo{person}{Craig
  Disselkoen}, \bibinfo{person}{Klaus~v. Gleissenthall}, \bibinfo{person}{Dean
  Tullsen}, \bibinfo{person}{Deian Stefan}, \bibinfo{person}{Tamara Rezk},
  {and} \bibinfo{person}{Gilles Barthe}.} \bibinfo{year}{2020}\natexlab{}.
\newblock \showarticletitle{Constant-Time Foundations for the New Spectre Era}.
  In \bibinfo{booktitle}{\emph{Proc. of the 41st ACM SIGPLAN Conf. on
  Programming Language Design and Implementation}} \emph{(\bibinfo{series}{PLDI
  2020})}. \bibinfo{publisher}{Association for Computing Machinery},
  \bibinfo{address}{New York, NY, USA}, \bibinfo{pages}{913--926}.
\newblock
\showISBNx{9781450376136}
\urldef\tempurl%
\url{https://doi.org/10.1145/3385412.3385970}
\showDOI{\tempurl}


\bibitem[\protect\citeauthoryear{Cauligi, Soeller, Johannesmeyer, Brown, Wahby,
  Renner, Gr\'{e}goire, Barthe, Jhala, and Stefan}{Cauligi
  et~al\mbox{.}}{2019}]%
        {FaCT-PLDI19}
\bibfield{author}{\bibinfo{person}{Sunjay Cauligi}, \bibinfo{person}{Gary
  Soeller}, \bibinfo{person}{Brian Johannesmeyer}, \bibinfo{person}{Fraser
  Brown}, \bibinfo{person}{Riad~S. Wahby}, \bibinfo{person}{John Renner},
  \bibinfo{person}{Benjamin Gr\'{e}goire}, \bibinfo{person}{Gilles Barthe},
  \bibinfo{person}{Ranjit Jhala}, {and} \bibinfo{person}{Deian Stefan}.}
  \bibinfo{year}{2019}\natexlab{}.
\newblock \showarticletitle{FaCT: A DSL for Timing-Sensitive Computation}. In
  \bibinfo{booktitle}{\emph{Proceedings of the 40th ACM SIGPLAN Conference on
  Programming Language Design and Implementation}} \emph{(\bibinfo{series}{PLDI
  2019})}. \bibinfo{publisher}{ACM}, \bibinfo{pages}{174--189}.
\newblock
\showISBNx{9781450367127}
\urldef\tempurl%
\url{https://doi.org/10.1145/3314221.3314605}
\showDOI{\tempurl}


\bibitem[\protect\citeauthoryear{Chevalier-Boisvert, Hendren, and
  Verbrugge}{Chevalier-Boisvert et~al\mbox{.}}{2010}]%
        {Boisvert-CC10}
\bibfield{author}{\bibinfo{person}{Maxime Chevalier-Boisvert},
  \bibinfo{person}{Laurie Hendren}, {and} \bibinfo{person}{Clark Verbrugge}.}
  \bibinfo{year}{2010}\natexlab{}.
\newblock \showarticletitle{Optimizing Matlab through Just-In-Time
  Specialization}. In \bibinfo{booktitle}{\emph{Compiler Construction}},
  \bibfield{editor}{\bibinfo{person}{Rajiv Gupta}} (Ed.).
  \bibinfo{publisher}{Springer Berlin Heidelberg}, \bibinfo{address}{Berlin,
  Heidelberg}, \bibinfo{pages}{46--65}.
\newblock
\showISBNx{978-3-642-11970-5}


\bibitem[\protect\citeauthoryear{Cleemput, Coppens, and De~Sutter}{Cleemput
  et~al\mbox{.}}{2012}]%
        {vanCleemput-TACO12}
\bibfield{author}{\bibinfo{person}{Jeroen~V. Cleemput}, \bibinfo{person}{Bart
  Coppens}, {and} \bibinfo{person}{Bjorn De~Sutter}.}
  \bibinfo{year}{2012}\natexlab{}.
\newblock \showarticletitle{Compiler Mitigations for Time Attacks on Modern X86
  Processors}.
\newblock \bibinfo{journal}{\emph{ACM Trans. Archit. Code Optim.}}
  \bibinfo{volume}{8}, \bibinfo{number}{4}, Article \bibinfo{articleno}{23}
  (\bibinfo{date}{Jan.} \bibinfo{year}{2012}), \bibinfo{numpages}{20}~pages.
\newblock
\showISSN{1544-3566}
\urldef\tempurl%
\url{https://doi.org/10.1145/2086696.2086702}
\showDOI{\tempurl}


\bibitem[\protect\citeauthoryear{Coppens, Verbauwhede, Bosschere, and
  Sutter}{Coppens et~al\mbox{.}}{2009}]%
        {Coppens-SP09}
\bibfield{author}{\bibinfo{person}{Bart Coppens}, \bibinfo{person}{Ingrid
  Verbauwhede}, \bibinfo{person}{Koen~De Bosschere}, {and}
  \bibinfo{person}{Bjorn~De Sutter}.} \bibinfo{year}{2009}\natexlab{}.
\newblock \showarticletitle{Practical Mitigations for Timing-Based Side-Channel
  Attacks on Modern X86 Processors}. In \bibinfo{booktitle}{\emph{Proceedings
  of the 2009 30th IEEE Symposium on Security and Privacy}}
  \emph{(\bibinfo{series}{SP '09})}. \bibinfo{publisher}{IEEE Computer
  Society}, \bibinfo{address}{USA}, \bibinfo{pages}{45--60}.
\newblock
\showISBNx{9780769536330}
\urldef\tempurl%
\url{https://doi.org/10.1109/SP.2009.19}
\showDOI{\tempurl}


\bibitem[\protect\citeauthoryear{Daniel, Bardin, and Rezk}{Daniel
  et~al\mbox{.}}{2020}]%
        {binsecrel-SP20}
\bibfield{author}{\bibinfo{person}{Lesly-Ann Daniel},
  \bibinfo{person}{S{\'e}bastien Bardin}, {and} \bibinfo{person}{Tamara Rezk}.}
  \bibinfo{year}{2020}\natexlab{}.
\newblock \showarticletitle{{Binsec/Rel}: Efficient Relational Symbolic
  Execution for Constant-Time at Binary-Level}. In
  \bibinfo{booktitle}{\emph{Proceedings of the 2020 IEEE Symposium on Security
  and Privacy}} \emph{(\bibinfo{series}{SP '20})}. \bibinfo{publisher}{IEEE
  Computer Society}.
\newblock


\bibitem[\protect\citeauthoryear{D'Elia, Coppa, Nicchi, Palmaro, and
  Cavallaro}{D'Elia et~al\mbox{.}}{2019}]%
        {SoK-DBI}
\bibfield{author}{\bibinfo{person}{Daniele~Cono D'Elia},
  \bibinfo{person}{Emilio Coppa}, \bibinfo{person}{Simone Nicchi},
  \bibinfo{person}{Federico Palmaro}, {and} \bibinfo{person}{Lorenzo
  Cavallaro}.} \bibinfo{year}{2019}\natexlab{}.
\newblock \showarticletitle{{SoK}: Using Dynamic Binary Instrumentation for
  Security (And How You May Get Caught Red Handed)}. In
  \bibinfo{booktitle}{\emph{Proc. of the 2019 ACM Asia Conference on Computer
  and Communications Security}} \emph{(\bibinfo{series}{Asia CCS '19})}.
  \bibinfo{publisher}{ACM}, \bibinfo{pages}{15--27}.
\newblock
\urldef\tempurl%
\url{https://doi.org/10.1145/3321705.3329819}
\showDOI{\tempurl}


\bibitem[\protect\citeauthoryear{D'Elia, Demetrescu, and Finocchi}{D'Elia
  et~al\mbox{.}}{2011}]%
        {Delia-PLDI11}
\bibfield{author}{\bibinfo{person}{Daniele~Cono D'Elia}, \bibinfo{person}{Camil
  Demetrescu}, {and} \bibinfo{person}{Irene Finocchi}.}
  \bibinfo{year}{2011}\natexlab{}.
\newblock \showarticletitle{Mining Hot Calling Contexts in Small Space}. In
  \bibinfo{booktitle}{\emph{Proceedings of the 32nd ACM SIGPLAN Conference on
  Programming Language Design and Implementation}} (San Jose, California, USA)
  \emph{(\bibinfo{series}{PLDI '11})}. \bibinfo{publisher}{Association for
  Computing Machinery}, \bibinfo{address}{New York, NY, USA},
  \bibinfo{pages}{516--527}.
\newblock
\showISBNx{9781450306638}
\urldef\tempurl%
\url{https://doi.org/10.1145/1993498.1993559}
\showDOI{\tempurl}


\bibitem[\protect\citeauthoryear{D'Elia, Demetrescu, and Finocchi}{D'Elia
  et~al\mbox{.}}{2016}]%
        {hcct-spe}
\bibfield{author}{\bibinfo{person}{Daniele~Cono D'Elia}, \bibinfo{person}{Camil
  Demetrescu}, {and} \bibinfo{person}{Irene Finocchi}.}
  \bibinfo{year}{2016}\natexlab{}.
\newblock \showarticletitle{Mining Hot Calling Contexts in Small Space}.
\newblock \bibinfo{journal}{\emph{Software: Practice and Experience}}
  \bibinfo{volume}{46}, \bibinfo{number}{8} (\bibinfo{year}{2016}),
  \bibinfo{pages}{1131--1152}.
\newblock
\showISSN{0038-0644}
\urldef\tempurl%
\url{https://doi.org/10.1002/spe.2348}
\showDOI{\tempurl}


\bibitem[\protect\citeauthoryear{Dinesh, Burow, Xu, and Payer}{Dinesh
  et~al\mbox{.}}{2020}]%
        {Retrowrite-SP20}
\bibfield{author}{\bibinfo{person}{Sushant Dinesh}, \bibinfo{person}{Nathan
  Burow}, \bibinfo{person}{Dongyan Xu}, {and} \bibinfo{person}{Mathias Payer}.}
  \bibinfo{year}{2020}\natexlab{}.
\newblock \showarticletitle{RetroWrite: Statically Instrumenting COTS Binaries
  for Fuzzing and Sanitization}.
\newblock  (\bibinfo{year}{2020}).
\newblock


\bibitem[\protect\citeauthoryear{Downs}{Downs}{2020}]%
        {hse}
\bibfield{author}{\bibinfo{person}{Travis Downs}.}
  \bibinfo{year}{2020}\natexlab{}.
\newblock \bibinfo{title}{Hardware Store Elimination}.
\newblock
  \bibinfo{howpublished}{\url{https://travisdowns.github.io/blog/2020/05/13/intel-zero-opt.html}}.
\newblock


\bibitem[\protect\citeauthoryear{Fioraldi, Maier, Ei{\ss}feldt, and
  Heuse}{Fioraldi et~al\mbox{.}}{2020}]%
        {aflpp-WOOT20}
\bibfield{author}{\bibinfo{person}{Andrea Fioraldi}, \bibinfo{person}{Dominik
  Maier}, \bibinfo{person}{Heiko Ei{\ss}feldt}, {and} \bibinfo{person}{Marc
  Heuse}.} \bibinfo{year}{2020}\natexlab{}.
\newblock \showarticletitle{AFL++ : Combining Incremental Steps of Fuzzing
  Research}. In \bibinfo{booktitle}{\emph{14th {USENIX} Workshop on Offensive
  Technologies ({WOOT} 20)}}. \bibinfo{publisher}{{USENIX} Association}.
\newblock


\bibitem[\protect\citeauthoryear{Fletcher, Dijk, and Devadas}{Fletcher
  et~al\mbox{.}}{2012}]%
        {ASCEND-STC12}
\bibfield{author}{\bibinfo{person}{Christopher~W. Fletcher},
  \bibinfo{person}{Marten~van Dijk}, {and} \bibinfo{person}{Srinivas Devadas}.}
  \bibinfo{year}{2012}\natexlab{}.
\newblock \showarticletitle{A Secure Processor Architecture for Encrypted
  Computation on Untrusted Programs}. In \bibinfo{booktitle}{\emph{Proceedings
  of the Seventh ACM Workshop on Scalable Trusted Computing}} (Raleigh, North
  Carolina, USA) \emph{(\bibinfo{series}{STC '12})}.
  \bibinfo{publisher}{Association for Computing Machinery},
  \bibinfo{address}{New York, NY, USA}, \bibinfo{pages}{3--8}.
\newblock
\showISBNx{9781450316620}
\urldef\tempurl%
\url{https://doi.org/10.1145/2382536.2382540}
\showDOI{\tempurl}


\bibitem[\protect\citeauthoryear{Fletchery, Ren, Yu, Van~Dijk, Khan, and
  Devadas}{Fletchery et~al\mbox{.}}{2014}]%
        {fletchery-hpca2014}
\bibfield{author}{\bibinfo{person}{Christopher~W Fletchery},
  \bibinfo{person}{Ling Ren}, \bibinfo{person}{Xiangyao Yu},
  \bibinfo{person}{Marten Van~Dijk}, \bibinfo{person}{Omer Khan}, {and}
  \bibinfo{person}{Srinivas Devadas}.} \bibinfo{year}{2014}\natexlab{}.
\newblock \showarticletitle{Suppressing the oblivious ram timing channel while
  making information leakage and program efficiency trade-offs}. In
  \bibinfo{booktitle}{\emph{2014 IEEE 20th International Symposium on High
  Performance Computer Architecture (HPCA)}}. IEEE, \bibinfo{pages}{213--224}.
\newblock


\bibitem[\protect\citeauthoryear{Gan, Zhang, Chen, Zhao, Qin, Wu, and Chen}{Gan
  et~al\mbox{.}}{2020}]%
        {GREYONE-SEC20}
\bibfield{author}{\bibinfo{person}{Shuitao Gan}, \bibinfo{person}{Chao Zhang},
  \bibinfo{person}{Peng Chen}, \bibinfo{person}{Bodong Zhao},
  \bibinfo{person}{Xiaojun Qin}, \bibinfo{person}{Dong Wu}, {and}
  \bibinfo{person}{Zuoning Chen}.} \bibinfo{year}{2020}\natexlab{}.
\newblock \showarticletitle{{GREYONE}: Data Flow Sensitive Fuzzing}. In
  \bibinfo{booktitle}{\emph{29th {USENIX} Security Symposium ({USENIX} Security
  20)}}. \bibinfo{publisher}{{USENIX} Association},
  \bibinfo{pages}{2577--2594}.
\newblock
\showISBNx{978-1-939133-17-5}
\urldef\tempurl%
\url{https://www.usenix.org/conference/usenixsecurity20/presentation/gan}
\showURL{%
\tempurl}


\bibitem[\protect\citeauthoryear{Goldreich and Ostrovsky}{Goldreich and
  Ostrovsky}{1996}]%
        {ORAM-JACM96}
\bibfield{author}{\bibinfo{person}{Oded Goldreich} {and}
  \bibinfo{person}{Rafail Ostrovsky}.} \bibinfo{year}{1996}\natexlab{}.
\newblock \showarticletitle{Software Protection and Simulation on Oblivious
  RAMs}.
\newblock \bibinfo{journal}{\emph{J. ACM}} \bibinfo{volume}{43},
  \bibinfo{number}{3} (\bibinfo{date}{May} \bibinfo{year}{1996}),
  \bibinfo{pages}{431?473}.
\newblock
\showISSN{0004-5411}
\urldef\tempurl%
\url{https://doi.org/10.1145/233551.233553}
\showDOI{\tempurl}


\bibitem[\protect\citeauthoryear{Gras, Giuffrida, Kurth, Bos, and Razavi}{Gras
  et~al\mbox{.}}{2020}]%
        {gras-ndss20}
\bibfield{author}{\bibinfo{person}{Ben Gras}, \bibinfo{person}{Cristiano
  Giuffrida}, \bibinfo{person}{Michael Kurth}, \bibinfo{person}{Herbert Bos},
  {and} \bibinfo{person}{Kaveh Razavi}.} \bibinfo{year}{2020}\natexlab{}.
\newblock \showarticletitle{ABSynthe: Automatic Blackbox Side-channel Synthesis
  on Commodity Microarchitectures}.
\newblock
\urldef\tempurl%
\url{https://doi.org/10.14722/ndss.2020.23018}
\showDOI{\tempurl}


\bibitem[\protect\citeauthoryear{Gruss, Lettner, Schuster, Ohrimenko, Haller,
  and Costa}{Gruss et~al\mbox{.}}{2017}]%
        {gruss-usenix17}
\bibfield{author}{\bibinfo{person}{Daniel Gruss}, \bibinfo{person}{Julian
  Lettner}, \bibinfo{person}{Felix Schuster}, \bibinfo{person}{Olga Ohrimenko},
  \bibinfo{person}{Istvan Haller}, {and} \bibinfo{person}{Manuel Costa}.}
  \bibinfo{year}{2017}\natexlab{}.
\newblock \showarticletitle{Strong and Efficient Cache Side-Channel Protection
  Using Hardware Transactional Memory}. In
  \bibinfo{booktitle}{\emph{Proceedings of the 26th USENIX Conference on
  Security Symposium}} \emph{(\bibinfo{series}{SEC'17})}.
  \bibinfo{publisher}{USENIX Association}, \bibinfo{address}{USA},
  \bibinfo{pages}{217--233}.
\newblock
\showISBNx{9781931971409}


\bibitem[\protect\citeauthoryear{{He}, {Emmi}, and {Ciocarlie}}{{He}
  et~al\mbox{.}}{2020}]%
        {ctfuzz}
\bibfield{author}{\bibinfo{person}{S. {He}}, \bibinfo{person}{M. {Emmi}}, {and}
  \bibinfo{person}{G. {Ciocarlie}}.} \bibinfo{year}{2020}\natexlab{}.
\newblock \showarticletitle{ct-fuzz: Fuzzing for Timing Leaks}. In
  \bibinfo{booktitle}{\emph{2020 IEEE 13th International Conference on Software
  Testing, Validation and Verification (ICST)}}. \bibinfo{pages}{466--471}.
\newblock
\urldef\tempurl%
\url{https://doi.org/10.1109/ICST46399.2020.00063}
\showDOI{\tempurl}


\bibitem[\protect\citeauthoryear{Hunger, Kazdagli, Rawat, Dimakis, Vishwanath,
  and Tiwari}{Hunger et~al\mbox{.}}{2015}]%
        {hunger-hpca2015}
\bibfield{author}{\bibinfo{person}{Casen Hunger}, \bibinfo{person}{Mikhail
  Kazdagli}, \bibinfo{person}{Ankit Rawat}, \bibinfo{person}{Alex Dimakis},
  \bibinfo{person}{Sriram Vishwanath}, {and} \bibinfo{person}{Mohit Tiwari}.}
  \bibinfo{year}{2015}\natexlab{}.
\newblock \showarticletitle{Understanding contention-based channels and using
  them for defense}. In \bibinfo{booktitle}{\emph{2015 IEEE 21st International
  Symposium on High Performance Computer Architecture (HPCA)}}. IEEE,
  \bibinfo{pages}{639--650}.
\newblock


\bibitem[\protect\citeauthoryear{Intel}{Intel}{2020}]%
        {intelstmt}
\bibfield{author}{\bibinfo{person}{Intel}.} \bibinfo{year}{2020}\natexlab{}.
\newblock \showarticletitle{Guidelines for Mitigating Timing Side Channels
  Against Cryptographic Implementations}.
\newblock \bibinfo{journal}{\emph{Developer Zone - Secure Coding}}
  (\bibinfo{year}{2020}).
\newblock
\urldef\tempurl%
\url{https://software.intel.com/security-software-guidance/secure-coding/guidelines-mitigating-timing-side-channels-against-cryptographic-implementations}
\showURL{%
\tempurl}


\bibitem[\protect\citeauthoryear{Kim, Peinado, and Mainar-Ruiz}{Kim
  et~al\mbox{.}}{2012}]%
        {stealthmem-usenix12}
\bibfield{author}{\bibinfo{person}{Taesoo Kim}, \bibinfo{person}{Marcus
  Peinado}, {and} \bibinfo{person}{Gloria Mainar-Ruiz}.}
  \bibinfo{year}{2012}\natexlab{}.
\newblock \showarticletitle{$\{$STEALTHMEM$\}$: System-level protection against
  cache-based side channel attacks in the cloud}. In
  \bibinfo{booktitle}{\emph{Presented as part of the 21st $\{$USENIX$\}$
  Security Symposium ($\{$USENIX$\}$ Security 12)}}. \bibinfo{pages}{189--204}.
\newblock


\bibitem[\protect\citeauthoryear{Kirsch, Jonischkeit, Kittel, Zarras, and
  Eckert}{Kirsch et~al\mbox{.}}{2017}]%
        {Kirsch-IFIPSEC17}
\bibfield{author}{\bibinfo{person}{Julian Kirsch}, \bibinfo{person}{Clemens
  Jonischkeit}, \bibinfo{person}{Thomas Kittel}, \bibinfo{person}{Apostolis
  Zarras}, {and} \bibinfo{person}{Claudia Eckert}.}
  \bibinfo{year}{2017}\natexlab{}.
\newblock \showarticletitle{Combating Control Flow Linearization}. In
  \bibinfo{booktitle}{\emph{ICT Systems Security and Privacy Protection}},
  \bibfield{editor}{\bibinfo{person}{Sabrina De~Capitani~di Vimercati} {and}
  \bibinfo{person}{Fabio Martinelli}} (Eds.). \bibinfo{publisher}{Springer
  International Publishing}, \bibinfo{address}{Cham},
  \bibinfo{pages}{385--398}.
\newblock
\showISBNx{978-3-319-58469-0}


\bibitem[\protect\citeauthoryear{{Kocher}, {Horn}, {Fogh}, {Genkin}, {Gruss},
  {Haas}, {Hamburg}, {Lipp}, {Mangard}, {Prescher}, {Schwarz}, and
  {Yarom}}{{Kocher} et~al\mbox{.}}{2019}]%
        {kocher-sp19}
\bibfield{author}{\bibinfo{person}{P. {Kocher}}, \bibinfo{person}{J. {Horn}},
  \bibinfo{person}{A. {Fogh}}, \bibinfo{person}{D. {Genkin}},
  \bibinfo{person}{D. {Gruss}}, \bibinfo{person}{W. {Haas}},
  \bibinfo{person}{M. {Hamburg}}, \bibinfo{person}{M. {Lipp}},
  \bibinfo{person}{S. {Mangard}}, \bibinfo{person}{T. {Prescher}},
  \bibinfo{person}{M. {Schwarz}}, {and} \bibinfo{person}{Y. {Yarom}}.}
  \bibinfo{year}{2019}\natexlab{}.
\newblock \showarticletitle{Spectre Attacks: Exploiting Speculative Execution}.
  In \bibinfo{booktitle}{\emph{2019 IEEE Symposium on Security and Privacy
  (SP)}}. \bibinfo{pages}{1--19}.
\newblock
\urldef\tempurl%
\url{https://doi.org/10.1109/SP.2019.00002}
\showDOI{\tempurl}


\bibitem[\protect\citeauthoryear{Lipp, Kogler, Oswald, Schwarz, Easdon,
  Canella, and Gruss}{Lipp et~al\mbox{.}}{2021}]%
        {Lipp2021Platypus}
\bibfield{author}{\bibinfo{person}{Moritz Lipp}, \bibinfo{person}{Andreas
  Kogler}, \bibinfo{person}{David Oswald}, \bibinfo{person}{Michael Schwarz},
  \bibinfo{person}{Catherine Easdon}, \bibinfo{person}{Claudio Canella}, {and}
  \bibinfo{person}{Daniel Gruss}.} \bibinfo{year}{2021}\natexlab{}.
\newblock \showarticletitle{{PLATYPUS: Software-based Power Side-Channel
  Attacks on x86}}. In \bibinfo{booktitle}{\emph{IEEE S\&P}}.
\newblock


\bibitem[\protect\citeauthoryear{Lipp, Schwarz, Gruss, Prescher, Haas, Fogh,
  Horn, Mangard, Kocher, Genkin, Yarom, and Hamburg}{Lipp
  et~al\mbox{.}}{2018}]%
        {lipp-usenix18}
\bibfield{author}{\bibinfo{person}{Moritz Lipp}, \bibinfo{person}{Michael
  Schwarz}, \bibinfo{person}{Daniel Gruss}, \bibinfo{person}{Thomas Prescher},
  \bibinfo{person}{Werner Haas}, \bibinfo{person}{Anders Fogh},
  \bibinfo{person}{Jann Horn}, \bibinfo{person}{Stefan Mangard},
  \bibinfo{person}{Paul Kocher}, \bibinfo{person}{Daniel Genkin},
  \bibinfo{person}{Yuval Yarom}, {and} \bibinfo{person}{Mike Hamburg}.}
  \bibinfo{year}{2018}\natexlab{}.
\newblock \showarticletitle{Meltdown: Reading Kernel Memory from User Space}.
  In \bibinfo{booktitle}{\emph{Proceedings of the 27th USENIX Conference on
  Security Symposium}} (Baltimore, MD, USA) \emph{(\bibinfo{series}{SEC'18})}.
  \bibinfo{publisher}{USENIX Association}, \bibinfo{address}{USA},
  \bibinfo{pages}{973–990}.
\newblock
\showISBNx{9781931971461}


\bibitem[\protect\citeauthoryear{Liu, Harris, Maas, Hicks, Tiwari, and Shi}{Liu
  et~al\mbox{.}}{2015}]%
        {GHOSTRIDER-ASPLOS15}
\bibfield{author}{\bibinfo{person}{Chang Liu}, \bibinfo{person}{Austin Harris},
  \bibinfo{person}{Martin Maas}, \bibinfo{person}{Michael Hicks},
  \bibinfo{person}{Mohit Tiwari}, {and} \bibinfo{person}{Elaine Shi}.}
  \bibinfo{year}{2015}\natexlab{}.
\newblock \showarticletitle{GhostRider: A Hardware-Software System for Memory
  Trace Oblivious Computation}. In \bibinfo{booktitle}{\emph{Proceedings of the
  Twentieth International Conference on Architectural Support for Programming
  Languages and Operating Systems}} (Istanbul, Turkey)
  \emph{(\bibinfo{series}{ASPLOS '15})}. \bibinfo{publisher}{Association for
  Computing Machinery}, \bibinfo{address}{New York, NY, USA},
  \bibinfo{pages}{87--101}.
\newblock
\showISBNx{9781450328357}
\urldef\tempurl%
\url{https://doi.org/10.1145/2694344.2694385}
\showDOI{\tempurl}


\bibitem[\protect\citeauthoryear{Liu, Hicks, and Shi}{Liu
  et~al\mbox{.}}{2013}]%
        {MTO-CSF13}
\bibfield{author}{\bibinfo{person}{Chang Liu}, \bibinfo{person}{Michael Hicks},
  {and} \bibinfo{person}{Elaine Shi}.} \bibinfo{year}{2013}\natexlab{}.
\newblock \showarticletitle{Memory Trace Oblivious Program Execution}. In
  \bibinfo{booktitle}{\emph{Proceedings of the 2013 IEEE 26th Computer Security
  Foundations Symposium}} \emph{(\bibinfo{series}{CSF '13})}.
  \bibinfo{publisher}{IEEE Computer Society}, \bibinfo{address}{USA},
  \bibinfo{pages}{51--65}.
\newblock
\showISBNx{9780769550312}
\urldef\tempurl%
\url{https://doi.org/10.1109/CSF.2013.11}
\showDOI{\tempurl}


\bibitem[\protect\citeauthoryear{Maas, Love, Stefanov, Tiwari, Shi, Asanovic,
  Kubiatowicz, and Song}{Maas et~al\mbox{.}}{2013}]%
        {PHANTOM-CCS13}
\bibfield{author}{\bibinfo{person}{Martin Maas}, \bibinfo{person}{Eric Love},
  \bibinfo{person}{Emil Stefanov}, \bibinfo{person}{Mohit Tiwari},
  \bibinfo{person}{Elaine Shi}, \bibinfo{person}{Krste Asanovic},
  \bibinfo{person}{John Kubiatowicz}, {and} \bibinfo{person}{Dawn Song}.}
  \bibinfo{year}{2013}\natexlab{}.
\newblock \showarticletitle{PHANTOM: Practical Oblivious Computation in a
  Secure Processor}. In \bibinfo{booktitle}{\emph{Proceedings of the 2013 ACM
  SIGSAC Conference on Computer \& Communications Security}}
  \emph{(\bibinfo{series}{CCS '13})}. \bibinfo{publisher}{Association for
  Computing Machinery}, \bibinfo{address}{New York, NY, USA},
  \bibinfo{pages}{311--324}.
\newblock
\showISBNx{9781450324779}
\urldef\tempurl%
\url{https://doi.org/10.1145/2508859.2516692}
\showDOI{\tempurl}


\bibitem[\protect\citeauthoryear{Mantel and Starostin}{Mantel and
  Starostin}{2015}]%
        {Mantel-ESORICS15}
\bibfield{author}{\bibinfo{person}{Heiko Mantel} {and} \bibinfo{person}{Artem
  Starostin}.} \bibinfo{year}{2015}\natexlab{}.
\newblock \showarticletitle{Transforming Out Timing Leaks, More or Less}. In
  \bibinfo{booktitle}{\emph{Proceedings, Part I, of the 20th European Symposium
  on Computer Security -- ESORICS 2015 - Volume 9326}}.
  \bibinfo{publisher}{Springer-Verlag}, \bibinfo{address}{Berlin, Heidelberg},
  \bibinfo{pages}{447--467}.
\newblock
\showISBNx{9783319241739}
\urldef\tempurl%
\url{https://doi.org/10.1007/978-3-319-24174-6_23}
\showDOI{\tempurl}


\bibitem[\protect\citeauthoryear{Martin, Demme, and Sethumadhavan}{Martin
  et~al\mbox{.}}{2012}]%
        {timewarp-isca12}
\bibfield{author}{\bibinfo{person}{Robert Martin}, \bibinfo{person}{John
  Demme}, {and} \bibinfo{person}{Simha Sethumadhavan}.}
  \bibinfo{year}{2012}\natexlab{}.
\newblock \showarticletitle{Timewarp: Rethinking timekeeping and performance
  monitoring mechanisms to mitigate side-channel attacks}. In
  \bibinfo{booktitle}{\emph{2012 39th Annual International Symposium on
  Computer Architecture (ISCA)}}. IEEE, \bibinfo{pages}{118--129}.
\newblock


\bibitem[\protect\citeauthoryear{Milanova, Rountev, and Ryder}{Milanova
  et~al\mbox{.}}{2002}]%
        {Milanova-ISSTA02}
\bibfield{author}{\bibinfo{person}{Ana Milanova}, \bibinfo{person}{Atanas
  Rountev}, {and} \bibinfo{person}{Barbara~G. Ryder}.}
  \bibinfo{year}{2002}\natexlab{}.
\newblock \showarticletitle{Parameterized Object Sensitivity for Points-to and
  Side-Effect Analyses for Java} \emph{(\bibinfo{series}{ISSTA '02})}.
  \bibinfo{publisher}{Association for Computing Machinery},
  \bibinfo{address}{New York, NY, USA}, \bibinfo{pages}{1--11}.
\newblock
\showISBNx{1581135629}
\urldef\tempurl%
\url{https://doi.org/10.1145/566172.566174}
\showDOI{\tempurl}


\bibitem[\protect\citeauthoryear{Moghimi, Wichelmann, Eisenbarth, and
  Sunar}{Moghimi et~al\mbox{.}}{2019}]%
        {MemJam}
\bibfield{author}{\bibinfo{person}{Ahmad Moghimi}, \bibinfo{person}{Jan
  Wichelmann}, \bibinfo{person}{Thomas Eisenbarth}, {and} \bibinfo{person}{Berk
  Sunar}.} \bibinfo{year}{2019}\natexlab{}.
\newblock \showarticletitle{MemJam: A False Dependency Attack Against
  Constant-Time Crypto Implementations}.
\newblock \bibinfo{journal}{\emph{Int. J. Parallel Program.}}
  \bibinfo{volume}{47}, \bibinfo{number}{4} (\bibinfo{date}{Aug.}
  \bibinfo{year}{2019}), \bibinfo{pages}{538?570}.
\newblock
\showISSN{0885-7458}
\urldef\tempurl%
\url{https://doi.org/10.1007/s10766-018-0611-9}
\showDOI{\tempurl}


\bibitem[\protect\citeauthoryear{Moghimi, Bulck, Heninger, Piessens, and
  Sunar}{Moghimi et~al\mbox{.}}{2020}]%
        {COPYCAT-SEC20}
\bibfield{author}{\bibinfo{person}{Daniel Moghimi}, \bibinfo{person}{Jo~Van
  Bulck}, \bibinfo{person}{Nadia Heninger}, \bibinfo{person}{Frank Piessens},
  {and} \bibinfo{person}{Berk Sunar}.} \bibinfo{year}{2020}\natexlab{}.
\newblock \showarticletitle{CopyCat: Controlled Instruction-Level Attacks on
  Enclaves}. In \bibinfo{booktitle}{\emph{29th {USENIX} Security Symposium
  ({USENIX} Security 20)}}. \bibinfo{publisher}{{USENIX} Association},
  \bibinfo{pages}{469--486}.
\newblock
\showISBNx{978-1-939133-17-5}
\urldef\tempurl%
\url{https://www.usenix.org/conference/usenixsecurity20/presentation/moghimi-copycat}
\showURL{%
\tempurl}


\bibitem[\protect\citeauthoryear{Molnar, Piotrowski, Schultz, and
  Wagner}{Molnar et~al\mbox{.}}{2005}]%
        {Molnar-ICISC05}
\bibfield{author}{\bibinfo{person}{David Molnar}, \bibinfo{person}{Matt
  Piotrowski}, \bibinfo{person}{David Schultz}, {and} \bibinfo{person}{David
  Wagner}.} \bibinfo{year}{2005}\natexlab{}.
\newblock \showarticletitle{The Program Counter Security Model: Automatic
  Detection and Removal of Control-Flow Side Channel Attacks}. In
  \bibinfo{booktitle}{\emph{Proceedings of the 8th International Conference on
  Information Security and Cryptology}} (Seoul, Korea)
  \emph{(\bibinfo{series}{ICISC'05})}. \bibinfo{publisher}{Springer-Verlag},
  \bibinfo{address}{Berlin, Heidelberg}, \bibinfo{pages}{156--168}.
\newblock
\urldef\tempurl%
\url{https://doi.org/10.1007/11734727\_14}
\showURL{%
\tempurl}


\bibitem[\protect\citeauthoryear{Muth, Watterson, and Debray}{Muth
  et~al\mbox{.}}{2000}]%
        {Muth-SAS2000}
\bibfield{author}{\bibinfo{person}{Robert Muth}, \bibinfo{person}{Scott
  Watterson}, {and} \bibinfo{person}{Saumya Debray}.}
  \bibinfo{year}{2000}\natexlab{}.
\newblock \showarticletitle{Code Specialization Based on Value Profiles}. In
  \bibinfo{booktitle}{\emph{Static Analysis}},
  \bibfield{editor}{\bibinfo{person}{Jens Palsberg}} (Ed.).
  \bibinfo{publisher}{Springer Berlin Heidelberg}, \bibinfo{address}{Berlin,
  Heidelberg}, \bibinfo{pages}{340--359}.
\newblock
\showISBNx{978-3-540-45099-3}


\bibitem[\protect\citeauthoryear{Osvik, Shamir, and Tromer}{Osvik
  et~al\mbox{.}}{2006}]%
        {osvik-rsa06}
\bibfield{author}{\bibinfo{person}{Dag~Arne Osvik}, \bibinfo{person}{Adi
  Shamir}, {and} \bibinfo{person}{Eran Tromer}.}
  \bibinfo{year}{2006}\natexlab{}.
\newblock \showarticletitle{Cache Attacks and Countermeasures: The Case of
  AES}. In \bibinfo{booktitle}{\emph{Proceedings of the 2006 The
  Cryptographers' Track at the RSA Conference on Topics in Cryptology}} (San
  Jose, CA) \emph{(\bibinfo{series}{CT-RSA'06})}.
  \bibinfo{publisher}{Springer-Verlag}, \bibinfo{address}{Berlin, Heidelberg},
  \bibinfo{pages}{1–20}.
\newblock
\showISBNx{3540310339}
\urldef\tempurl%
\url{https://doi.org/10.1007/11605805_1}
\showDOI{\tempurl}


\bibitem[\protect\citeauthoryear{Rane, Lin, and Tiwari}{Rane
  et~al\mbox{.}}{2015}]%
        {RACCOON-SEC15}
\bibfield{author}{\bibinfo{person}{Ashay Rane}, \bibinfo{person}{Calvin Lin},
  {and} \bibinfo{person}{Mohit Tiwari}.} \bibinfo{year}{2015}\natexlab{}.
\newblock \showarticletitle{Raccoon: Closing Digital Side-Channels through
  Obfuscated Execution}. In \bibinfo{booktitle}{\emph{Proceedings of the 24th
  USENIX Conference on Security Symposium}} (Washington, D.C.)
  \emph{(\bibinfo{series}{SEC'15})}. \bibinfo{publisher}{USENIX Association},
  \bibinfo{address}{USA}, \bibinfo{pages}{431--446}.
\newblock
\showISBNx{9781931971232}


\bibitem[\protect\citeauthoryear{Recoules, Bardin, Bonichon, Mounier, and
  Potet}{Recoules et~al\mbox{.}}{2019}]%
        {bardin-ase19}
\bibfield{author}{\bibinfo{person}{Fr\'{e}d\'{e}ric Recoules},
  \bibinfo{person}{S\'{e}bastien Bardin}, \bibinfo{person}{Richard Bonichon},
  \bibinfo{person}{Laurent Mounier}, {and} \bibinfo{person}{Marie-Laure
  Potet}.} \bibinfo{year}{2019}\natexlab{}.
\newblock \showarticletitle{Get Rid of Inline Assembly through
  Verification-Oriented Lifting}. In \bibinfo{booktitle}{\emph{Proceedings of
  the 34th IEEE/ACM International Conference on Automated Software
  Engineering}} (San Diego, California) \emph{(\bibinfo{series}{ASE '19})}.
  \bibinfo{publisher}{IEEE Press}, \bibinfo{pages}{577–589}.
\newblock
\showISBNx{9781728125084}
\urldef\tempurl%
\url{https://doi.org/10.1109/ASE.2019.00060}
\showDOI{\tempurl}


\bibitem[\protect\citeauthoryear{Rodrigues, Quint\~{a}o Pereira, and
  Aranha}{Rodrigues et~al\mbox{.}}{2016}]%
        {FlowTracker-CC16}
\bibfield{author}{\bibinfo{person}{Bruno Rodrigues},
  \bibinfo{person}{Fernando~Magno Quint\~{a}o Pereira}, {and}
  \bibinfo{person}{Diego~F. Aranha}.} \bibinfo{year}{2016}\natexlab{}.
\newblock \showarticletitle{Sparse Representation of Implicit Flows with
  Applications to Side-Channel Detection}. In
  \bibinfo{booktitle}{\emph{Proceedings of the 25th International Conference on
  Compiler Construction}} (Barcelona, Spain) \emph{(\bibinfo{series}{CC
  2016})}. \bibinfo{publisher}{Association for Computing Machinery},
  \bibinfo{address}{New York, NY, USA}, \bibinfo{pages}{110--120}.
\newblock
\showISBNx{9781450342414}
\urldef\tempurl%
\url{https://doi.org/10.1145/2892208.2892230}
\showDOI{\tempurl}


\bibitem[\protect\citeauthoryear{Rosen, Wegman, and Zadeck}{Rosen
  et~al\mbox{.}}{1988}]%
        {Rosen-POPL88}
\bibfield{author}{\bibinfo{person}{B.~K. Rosen}, \bibinfo{person}{M.~N.
  Wegman}, {and} \bibinfo{person}{F.~K. Zadeck}.}
  \bibinfo{year}{1988}\natexlab{}.
\newblock \showarticletitle{Global Value Numbers and Redundant Computations}.
  In \bibinfo{booktitle}{\emph{Proceedings of the 15th ACM SIGPLAN-SIGACT
  Symposium on Principles of Programming Languages}} (San Diego, California,
  USA) \emph{(\bibinfo{series}{POPL '88})}. \bibinfo{publisher}{Association for
  Computing Machinery}, \bibinfo{address}{New York, NY, USA},
  \bibinfo{pages}{12--27}.
\newblock
\showISBNx{0897912527}
\urldef\tempurl%
\url{https://doi.org/10.1145/73560.73562}
\showDOI{\tempurl}


\bibitem[\protect\citeauthoryear{Schwarz, Lipp, Moghimi, Van~Bulck, Stecklina,
  Prescher, and Gruss}{Schwarz et~al\mbox{.}}{2019}]%
        {Schwarz-ccs19}
\bibfield{author}{\bibinfo{person}{Michael Schwarz}, \bibinfo{person}{Moritz
  Lipp}, \bibinfo{person}{Daniel Moghimi}, \bibinfo{person}{Jo Van~Bulck},
  \bibinfo{person}{Julian Stecklina}, \bibinfo{person}{Thomas Prescher}, {and}
  \bibinfo{person}{Daniel Gruss}.} \bibinfo{year}{2019}\natexlab{}.
\newblock \showarticletitle{ZombieLoad: Cross-Privilege-Boundary Data
  Sampling}. In \bibinfo{booktitle}{\emph{Proceedings of the 2019 ACM SIGSAC
  Conference on Computer and Communications Security}} (London, United Kingdom)
  \emph{(\bibinfo{series}{CCS '19})}. \bibinfo{publisher}{Association for
  Computing Machinery}, \bibinfo{address}{New York, NY, USA},
  \bibinfo{pages}{753–768}.
\newblock
\showISBNx{9781450367479}
\urldef\tempurl%
\url{https://doi.org/10.1145/3319535.3354252}
\showDOI{\tempurl}


\bibitem[\protect\citeauthoryear{Schwarzl, Canella, Gruss, and
  Schwarz}{Schwarzl et~al\mbox{.}}{2021}]%
        {Schwarzl2021Specfuscator}
\bibfield{author}{\bibinfo{person}{Martin Schwarzl}, \bibinfo{person}{Claudio
  Canella}, \bibinfo{person}{Daniel Gruss}, {and} \bibinfo{person}{Michael
  Schwarz}.} \bibinfo{year}{2021}\natexlab{}.
\newblock \showarticletitle{{Specfuscator: Evaluating Branch Removal as a
  Spectre Mitigation}}. In \bibinfo{booktitle}{\emph{Financial Cryptography and
  Data Security 2021}}.
\newblock


\bibitem[\protect\citeauthoryear{Shi, Chan, Stefanov, and Li}{Shi
  et~al\mbox{.}}{2011}]%
        {Shi-asiacrypt2011}
\bibfield{author}{\bibinfo{person}{Elaine Shi}, \bibinfo{person}{T.~H.~Hubert
  Chan}, \bibinfo{person}{Emil Stefanov}, {and} \bibinfo{person}{Mingfei Li}.}
  \bibinfo{year}{2011}\natexlab{}.
\newblock \showarticletitle{Oblivious {RAM} with O(({logN})3) Worst-Case Cost}.
\newblock In \bibinfo{booktitle}{\emph{In: Lee D.H., Wang X. (eds) Advances in
  Cryptology -- ASIACRYPT 2011. Lecture Notes in Computer Science, vol 7073}}.
  \bibinfo{publisher}{Springer Berlin Heidelberg}, \bibinfo{pages}{197--214}.
\newblock
\urldef\tempurl%
\url{https://doi.org/10.1007/978-3-642-25385-0_11}
\showDOI{\tempurl}


\bibitem[\protect\citeauthoryear{Smaragdakis and Balatsouras}{Smaragdakis and
  Balatsouras}{2015}]%
        {pointer-analysis-tutorial}
\bibfield{author}{\bibinfo{person}{Yannis Smaragdakis} {and}
  \bibinfo{person}{George Balatsouras}.} \bibinfo{year}{2015}\natexlab{}.
\newblock \showarticletitle{Pointer Analysis}.
\newblock \bibinfo{journal}{\emph{Found. and Trends in Prog. Lang.}}
  \bibinfo{volume}{2}, \bibinfo{number}{1} (\bibinfo{year}{2015}),
  \bibinfo{pages}{1--69}.
\newblock
\showISSN{2325-1107}
\urldef\tempurl%
\url{https://doi.org/10.1561/2500000014}
\showDOI{\tempurl}


\bibitem[\protect\citeauthoryear{Smaragdakis, Bravenboer, and
  Lhot\'{a}k}{Smaragdakis et~al\mbox{.}}{2011}]%
        {Smaragdakis-POPL11}
\bibfield{author}{\bibinfo{person}{Yannis Smaragdakis}, \bibinfo{person}{Martin
  Bravenboer}, {and} \bibinfo{person}{Ondrej Lhot\'{a}k}.}
  \bibinfo{year}{2011}\natexlab{}.
\newblock \showarticletitle{Pick Your Contexts Well: Understanding
  Object-Sensitivity}. In \bibinfo{booktitle}{\emph{Proceedings of the 38th
  Annual ACM SIGPLAN-SIGACT Symposium on Principles of Programming Languages}}
  (Austin, Texas, USA) \emph{(\bibinfo{series}{POPL '11})}.
  \bibinfo{publisher}{Association for Computing Machinery},
  \bibinfo{address}{New York, NY, USA}, \bibinfo{pages}{17--30}.
\newblock
\showISBNx{9781450304900}
\urldef\tempurl%
\url{https://doi.org/10.1145/1926385.1926390}
\showDOI{\tempurl}


\bibitem[\protect\citeauthoryear{Soares and Pereira}{Soares and
  Pereira}{2021}]%
        {Soares-CGO21}
\bibfield{author}{\bibinfo{person}{Luigi Soares} {and}
  \bibinfo{person}{Fernando Magno~Quintao Pereira}.}
  \bibinfo{year}{2021}\natexlab{}.
\newblock \showarticletitle{Memory-Safe Elimination of Side Channels}. In
  \bibinfo{booktitle}{\emph{(to appear) In Proceedings of the 2021 IEEE/ACM
  International Symposium on Code Generation and Optimization}}
  \emph{(\bibinfo{series}{CGO 2021})}.
\newblock


\bibitem[\protect\citeauthoryear{Somorovsky}{Somorovsky}{2016}]%
        {Somorovsky-CCS16}
\bibfield{author}{\bibinfo{person}{Juraj Somorovsky}.}
  \bibinfo{year}{2016}\natexlab{}.
\newblock \showarticletitle{Systematic Fuzzing and Testing of TLS Libraries}.
  In \bibinfo{booktitle}{\emph{Proceedings of the 2016 ACM SIGSAC Conference on
  Computer and Communications Security}} (Vienna, Austria)
  \emph{(\bibinfo{series}{CCS '16})}. \bibinfo{publisher}{Association for
  Computing Machinery}, \bibinfo{address}{New York, NY, USA},
  \bibinfo{pages}{1492--1504}.
\newblock
\showISBNx{9781450341394}
\urldef\tempurl%
\url{https://doi.org/10.1145/2976749.2978411}
\showDOI{\tempurl}


\bibitem[\protect\citeauthoryear{Steensgaard}{Steensgaard}{1996}]%
        {Steensgaard-POPL96}
\bibfield{author}{\bibinfo{person}{Bjarne Steensgaard}.}
  \bibinfo{year}{1996}\natexlab{}.
\newblock \showarticletitle{Points-to Analysis in Almost Linear Time}. In
  \bibinfo{booktitle}{\emph{Proceedings of the 23rd ACM SIGPLAN-SIGACT
  Symposium on Principles of Programming Languages}} (St. Petersburg Beach,
  Florida, USA) \emph{(\bibinfo{series}{POPL '96})}.
  \bibinfo{publisher}{Association for Computing Machinery},
  \bibinfo{address}{New York, NY, USA}, \bibinfo{pages}{32--41}.
\newblock
\showISBNx{0897917693}
\urldef\tempurl%
\url{https://doi.org/10.1145/237721.237727}
\showDOI{\tempurl}


\bibitem[\protect\citeauthoryear{Stefanov, van Dijk, Shi, Fletcher, Ren, Yu,
  and Devadas}{Stefanov et~al\mbox{.}}{2013}]%
        {PATHORAM-CCS13}
\bibfield{author}{\bibinfo{person}{Emil Stefanov}, \bibinfo{person}{Marten van
  Dijk}, \bibinfo{person}{Elaine Shi}, \bibinfo{person}{Christopher Fletcher},
  \bibinfo{person}{Ling Ren}, \bibinfo{person}{Xiangyao Yu}, {and}
  \bibinfo{person}{Srinivas Devadas}.} \bibinfo{year}{2013}\natexlab{}.
\newblock \showarticletitle{Path ORAM: An Extremely Simple Oblivious RAM
  Protocol}. In \bibinfo{booktitle}{\emph{Proceedings of the 2013 ACM SIGSAC
  Conference on Computer \& Communications Security}}
  \emph{(\bibinfo{series}{CCS '13})}. \bibinfo{publisher}{Association for
  Computing Machinery}, \bibinfo{address}{New York, NY, USA},
  \bibinfo{pages}{299--310}.
\newblock
\showISBNx{9781450324779}
\urldef\tempurl%
\url{https://doi.org/10.1145/2508859.2516660}
\showDOI{\tempurl}


\bibitem[\protect\citeauthoryear{{Stephens}, {Biles}, {Boettcher}, {Eapen},
  {Eyole}, {Gabrielli}, {Horsnell}, {Magklis}, {Martinez}, {Premillieu},
  {Reid}, {Rico}, and {Walker}}{{Stephens} et~al\mbox{.}}{2017}]%
        {ARMSVE-MICRO17}
\bibfield{author}{\bibinfo{person}{N. {Stephens}}, \bibinfo{person}{S.
  {Biles}}, \bibinfo{person}{M. {Boettcher}}, \bibinfo{person}{J. {Eapen}},
  \bibinfo{person}{M. {Eyole}}, \bibinfo{person}{G. {Gabrielli}},
  \bibinfo{person}{M. {Horsnell}}, \bibinfo{person}{G. {Magklis}},
  \bibinfo{person}{A. {Martinez}}, \bibinfo{person}{N. {Premillieu}},
  \bibinfo{person}{A. {Reid}}, \bibinfo{person}{A. {Rico}}, {and}
  \bibinfo{person}{P. {Walker}}.} \bibinfo{year}{2017}\natexlab{}.
\newblock \showarticletitle{The ARM Scalable Vector Extension}.
\newblock \bibinfo{journal}{\emph{IEEE Micro}} \bibinfo{volume}{37},
  \bibinfo{number}{2} (\bibinfo{year}{2017}), \bibinfo{pages}{26--39}.
\newblock
\urldef\tempurl%
\url{https://doi.org/10.1109/MM.2017.35}
\showDOI{\tempurl}


\bibitem[\protect\citeauthoryear{Suh, Clarke, Gassend, Van~Dijk, and
  Devadas}{Suh et~al\mbox{.}}{2003}]%
        {aegis-sc2003}
\bibfield{author}{\bibinfo{person}{G~Edward Suh}, \bibinfo{person}{Dwaine
  Clarke}, \bibinfo{person}{Blaise Gassend}, \bibinfo{person}{Marten Van~Dijk},
  {and} \bibinfo{person}{Srinivas Devadas}.} \bibinfo{year}{2003}\natexlab{}.
\newblock \showarticletitle{AEGIS: architecture for tamper-evident and
  tamper-resistant processing}. In \bibinfo{booktitle}{\emph{ACM International
  Conference on Supercomputing 25th Anniversary Volume}}.
  \bibinfo{pages}{357--368}.
\newblock


\bibitem[\protect\citeauthoryear{Sui and Xue}{Sui and Xue}{2016}]%
        {SVF-CC16}
\bibfield{author}{\bibinfo{person}{Yulei Sui} {and} \bibinfo{person}{Jingling
  Xue}.} \bibinfo{year}{2016}\natexlab{}.
\newblock \showarticletitle{SVF: Interprocedural Static Value-Flow Analysis in
  LLVM}. In \bibinfo{booktitle}{\emph{Proceedings of the 25th International
  Conference on Compiler Construction}} (Barcelona, Spain)
  \emph{(\bibinfo{series}{CC 2016})}. \bibinfo{publisher}{Association for
  Computing Machinery}, \bibinfo{address}{New York, NY, USA},
  \bibinfo{pages}{265--266}.
\newblock
\showISBNx{9781450342414}
\urldef\tempurl%
\url{https://doi.org/10.1145/2892208.2892235}
\showDOI{\tempurl}


\bibitem[\protect\citeauthoryear{{U.S. National Security Agency}}{{U.S.
  National Security Agency}}{2016}]%
        {nsa-faq}
\bibfield{author}{\bibinfo{person}{{U.S. National Security Agency}}.}
  \bibinfo{year}{2016}\natexlab{}.
\newblock \showarticletitle{Commercial National Security Algorithm Suite and
  Quantum Computing {FAQ}}.
\newblock  (\bibinfo{date}{Jan.} \bibinfo{year}{2016}).
\newblock


\bibitem[\protect\citeauthoryear{Van~Bulck, Minkin, Weisse, Genkin, Kasikci,
  Piessens, Silberstein, Wenisch, Yarom, and Strackx}{Van~Bulck
  et~al\mbox{.}}{2018}]%
        {bulk-usenix18}
\bibfield{author}{\bibinfo{person}{Jo Van~Bulck}, \bibinfo{person}{Marina
  Minkin}, \bibinfo{person}{Ofir Weisse}, \bibinfo{person}{Daniel Genkin},
  \bibinfo{person}{Baris Kasikci}, \bibinfo{person}{Frank Piessens},
  \bibinfo{person}{Mark Silberstein}, \bibinfo{person}{Thomas~F. Wenisch},
  \bibinfo{person}{Yuval Yarom}, {and} \bibinfo{person}{Raoul Strackx}.}
  \bibinfo{year}{2018}\natexlab{}.
\newblock \showarticletitle{Foreshadow: Extracting the Keys to the Intel SGX
  Kingdom with Transient out-of-Order Execution}. In
  \bibinfo{booktitle}{\emph{Proceedings of the 27th USENIX Conference on
  Security Symposium}} (Baltimore, MD, USA) \emph{(\bibinfo{series}{SEC'18})}.
  \bibinfo{publisher}{USENIX Association}, \bibinfo{address}{USA},
  \bibinfo{pages}{991–1008}.
\newblock
\showISBNx{9781931971461}


\bibitem[\protect\citeauthoryear{van~der Veen, Fratantonio, Lindorfer, Gruss,
  Maurice, Vigna, Bos, Razavi, and Giuffrida}{van~der Veen
  et~al\mbox{.}}{2016}]%
        {van_der_veen_drammer_2016}
\bibfield{author}{\bibinfo{person}{Victor van~der Veen},
  \bibinfo{person}{Yanick Fratantonio}, \bibinfo{person}{Martina Lindorfer},
  \bibinfo{person}{Daniel Gruss}, \bibinfo{person}{Clementine Maurice},
  \bibinfo{person}{Giovanni Vigna}, \bibinfo{person}{Herbert Bos},
  \bibinfo{person}{Kaveh Razavi}, {and} \bibinfo{person}{Cristiano Giuffrida}.}
  \bibinfo{year}{2016}\natexlab{}.
\newblock \showarticletitle{Drammer: Deterministic Rowhammer Attacks on Mobile
  Platforms}. In \bibinfo{booktitle}{\emph{Proceedings of the 2016 ACM SIGSAC
  Conference on Computer and Communications Security}} (Vienna, Austria)
  \emph{(\bibinfo{series}{CCS '16})}. \bibinfo{publisher}{Association for
  Computing Machinery}, \bibinfo{address}{New York, NY, USA},
  \bibinfo{pages}{1675--1689}.
\newblock
\showISBNx{9781450341394}
\urldef\tempurl%
\url{https://doi.org/10.1145/2976749.2978406}
\showDOI{\tempurl}


\bibitem[\protect\citeauthoryear{{van Schaik}, {Milburn}, {Österlund},
  {Frigo}, {Maisuradze}, {Razavi}, {Bos}, and {Giuffrida}}{{van Schaik}
  et~al\mbox{.}}{2019}]%
        {VanSchaik-sp19}
\bibfield{author}{\bibinfo{person}{S. {van Schaik}}, \bibinfo{person}{A.
  {Milburn}}, \bibinfo{person}{S. {Österlund}}, \bibinfo{person}{P. {Frigo}},
  \bibinfo{person}{G. {Maisuradze}}, \bibinfo{person}{K. {Razavi}},
  \bibinfo{person}{H. {Bos}}, {and} \bibinfo{person}{C. {Giuffrida}}.}
  \bibinfo{year}{2019}\natexlab{}.
\newblock \showarticletitle{RIDL: Rogue In-Flight Data Load}. In
  \bibinfo{booktitle}{\emph{2019 IEEE Symposium on Security and Privacy (SP)}}.
  \bibinfo{pages}{88--105}.
\newblock
\urldef\tempurl%
\url{https://doi.org/10.1109/SP.2019.00087}
\showDOI{\tempurl}


\bibitem[\protect\citeauthoryear{Vattikonda, Das, and Shacham}{Vattikonda
  et~al\mbox{.}}{2011}]%
        {vattikonda-wccs2011}
\bibfield{author}{\bibinfo{person}{Bhanu~C Vattikonda}, \bibinfo{person}{Sambit
  Das}, {and} \bibinfo{person}{Hovav Shacham}.}
  \bibinfo{year}{2011}\natexlab{}.
\newblock \showarticletitle{Eliminating fine grained timers in Xen}. In
  \bibinfo{booktitle}{\emph{Proceedings of the 3rd ACM workshop on Cloud
  computing security workshop}}. \bibinfo{pages}{41--46}.
\newblock


\bibitem[\protect\citeauthoryear{Wang, Zeldovich, Kaashoek, and
  Solar-Lezama}{Wang et~al\mbox{.}}{2013}]%
        {wang2013towards}
\bibfield{author}{\bibinfo{person}{Xi Wang}, \bibinfo{person}{Nickolai
  Zeldovich}, \bibinfo{person}{M.~Frans Kaashoek}, {and}
  \bibinfo{person}{Armando Solar-Lezama}.} \bibinfo{year}{2013}\natexlab{}.
\newblock \showarticletitle{Towards Optimization-Safe Systems: Analyzing the
  Impact of Undefined Behavior}. In \bibinfo{booktitle}{\emph{Proceedings of
  the Twenty-Fourth ACM Symposium on Operating Systems Principles}}
  \emph{(\bibinfo{series}{SOSP '13})}. \bibinfo{publisher}{Association for
  Computing Machinery}, \bibinfo{address}{New York, NY, USA},
  \bibinfo{pages}{260--275}.
\newblock
\showISBNx{9781450323888}
\urldef\tempurl%
\url{https://doi.org/10.1145/2517349.2522728}
\showDOI{\tempurl}


\bibitem[\protect\citeauthoryear{Wang and Lee}{Wang and Lee}{2007a}]%
        {wang-isca2007}
\bibfield{author}{\bibinfo{person}{Zhenghong Wang} {and}
  \bibinfo{person}{Ruby~B Lee}.} \bibinfo{year}{2007}\natexlab{a}.
\newblock \showarticletitle{New cache designs for thwarting software
  cache-based side channel attacks}. In \bibinfo{booktitle}{\emph{Proceedings
  of the 34th annual international symposium on Computer architecture (ISCA)}}.
  \bibinfo{pages}{494--505}.
\newblock


\bibitem[\protect\citeauthoryear{Wang and Lee}{Wang and Lee}{2007b}]%
        {Wang-ISCA07}
\bibfield{author}{\bibinfo{person}{Zhenghong Wang} {and}
  \bibinfo{person}{Ruby~B. Lee}.} \bibinfo{year}{2007}\natexlab{b}.
\newblock \showarticletitle{New Cache Designs for Thwarting Software
  Cache-Based Side Channel Attacks}. In \bibinfo{booktitle}{\emph{Proceedings
  of the 34th Annual International Symposium on Computer Architecture}} (San
  Diego, California, USA) \emph{(\bibinfo{series}{ISCA '07})}.
  \bibinfo{publisher}{Association for Computing Machinery},
  \bibinfo{address}{New York, NY, USA}, \bibinfo{pages}{494--505}.
\newblock
\showISBNx{9781595937063}
\urldef\tempurl%
\url{https://doi.org/10.1145/1250662.1250723}
\showDOI{\tempurl}


\bibitem[\protect\citeauthoryear{Wang and Lee}{Wang and Lee}{2008}]%
        {wang-microarchitecture2008}
\bibfield{author}{\bibinfo{person}{Zhenghong Wang} {and}
  \bibinfo{person}{Ruby~B Lee}.} \bibinfo{year}{2008}\natexlab{}.
\newblock \showarticletitle{A novel cache architecture with enhanced
  performance and security}. In \bibinfo{booktitle}{\emph{2008 41st IEEE/ACM
  International Symposium on Microarchitecture}}. IEEE,
  \bibinfo{pages}{83--93}.
\newblock


\bibitem[\protect\citeauthoryear{Weiser, Zankl, Spreitzer, Miller, Mangard, and
  Sigl}{Weiser et~al\mbox{.}}{2018}]%
        {data-usenix18}
\bibfield{author}{\bibinfo{person}{Samuel Weiser}, \bibinfo{person}{Andreas
  Zankl}, \bibinfo{person}{Raphael Spreitzer}, \bibinfo{person}{Katja Miller},
  \bibinfo{person}{Stefan Mangard}, {and} \bibinfo{person}{Georg Sigl}.}
  \bibinfo{year}{2018}\natexlab{}.
\newblock \showarticletitle{{DATA} {\textendash} Differential Address Trace
  Analysis: Finding Address-based Side-Channels in Binaries}. In
  \bibinfo{booktitle}{\emph{27th {USENIX} Security Symposium ({USENIX} Security
  18)}}. \bibinfo{publisher}{{USENIX} Association},
  \bibinfo{address}{Baltimore, MD}, \bibinfo{pages}{603--620}.
\newblock
\showISBNx{978-1-939133-04-5}
\urldef\tempurl%
\url{https://www.usenix.org/conference/usenixsecurity18/presentation/weiser}
\showURL{%
\tempurl}


\bibitem[\protect\citeauthoryear{Whaley and Lam}{Whaley and Lam}{2004}]%
        {Whaley-PLDI04}
\bibfield{author}{\bibinfo{person}{John Whaley} {and}
  \bibinfo{person}{Monica~S. Lam}.} \bibinfo{year}{2004}\natexlab{}.
\newblock \showarticletitle{Cloning-Based Context-Sensitive Pointer Alias
  Analysis Using Binary Decision Diagrams}. In
  \bibinfo{booktitle}{\emph{Proceedings of the ACM SIGPLAN 2004 Conference on
  Programming Language Design and Implementation}} (Washington DC, USA)
  \emph{(\bibinfo{series}{PLDI '04})}. \bibinfo{publisher}{Association for
  Computing Machinery}, \bibinfo{address}{New York, NY, USA},
  \bibinfo{pages}{131--144}.
\newblock
\showISBNx{1581138075}
\urldef\tempurl%
\url{https://doi.org/10.1145/996841.996859}
\showDOI{\tempurl}


\bibitem[\protect\citeauthoryear{Wu, Guo, Schaumont, and Wang}{Wu
  et~al\mbox{.}}{2018}]%
        {Wu-ISSTA18}
\bibfield{author}{\bibinfo{person}{Meng Wu}, \bibinfo{person}{Shengjian Guo},
  \bibinfo{person}{Patrick Schaumont}, {and} \bibinfo{person}{Chao Wang}.}
  \bibinfo{year}{2018}\natexlab{}.
\newblock \showarticletitle{Eliminating Timing Side-Channel Leaks Using Program
  Repair}. In \bibinfo{booktitle}{\emph{Proc. of the 27th ACM SIGSOFT Int.
  Symposium on Software Testing and Analysis}} \emph{(\bibinfo{series}{ISSTA
  2018})}. \bibinfo{publisher}{Association for Computing Machinery},
  \bibinfo{pages}{15--26}.
\newblock
\showISBNx{9781450356992}
\urldef\tempurl%
\url{https://doi.org/10.1145/3213846.3213851}
\showDOI{\tempurl}


\bibitem[\protect\citeauthoryear{Yarom and Falkner}{Yarom and Falkner}{2014}]%
        {yarom-usenix14}
\bibfield{author}{\bibinfo{person}{Yuval Yarom} {and} \bibinfo{person}{Katrina
  Falkner}.} \bibinfo{year}{2014}\natexlab{}.
\newblock \showarticletitle{FLUSH+RELOAD: A High Resolution, Low Noise, L3
  Cache Side-Channel Attack}. In \bibinfo{booktitle}{\emph{Proc. of the 23rd
  USENIX Security Symposium}} (San Diego, CA)
  \emph{(\bibinfo{series}{SEC'14})}. \bibinfo{publisher}{USENIX Association},
  \bibinfo{address}{USA}, \bibinfo{pages}{719--732}.
\newblock
\showISBNx{9781931971157}


\bibitem[\protect\citeauthoryear{Yarom, Genkin, and Heninger}{Yarom
  et~al\mbox{.}}{2016}]%
        {CacheBleed-CHES16}
\bibfield{author}{\bibinfo{person}{Yuval Yarom}, \bibinfo{person}{Daniel
  Genkin}, {and} \bibinfo{person}{Nadia Heninger}.}
  \bibinfo{year}{2016}\natexlab{}.
\newblock \showarticletitle{CacheBleed: A Timing Attack on OpenSSL Constant
  Time RSA}. In \bibinfo{booktitle}{\emph{Cryptographic Hardware and Embedded
  Systems -- CHES 2016}}, \bibfield{editor}{\bibinfo{person}{Benedikt
  Gierlichs} {and} \bibinfo{person}{Axel~Y. Poschmann}} (Eds.).
  \bibinfo{publisher}{Springer Berlin Heidelberg}, \bibinfo{address}{Berlin,
  Heidelberg}, \bibinfo{pages}{346--367}.
\newblock
\showISBNx{978-3-662-53140-2}


\bibitem[\protect\citeauthoryear{Yu and Kaser}{Yu and Kaser}{1997}]%
        {scc-toplas97}
\bibfield{author}{\bibinfo{person}{Ting Yu} {and} \bibinfo{person}{Owen
  Kaser}.} \bibinfo{year}{1997}\natexlab{}.
\newblock \showarticletitle{A Note on ''On the Conversion of Indirect to Direct
  Recursion''}.
\newblock \bibinfo{journal}{\emph{ACM Trans. Program. Lang. Syst.}}
  \bibinfo{volume}{19}, \bibinfo{number}{6} (\bibinfo{date}{Nov.}
  \bibinfo{year}{1997}), \bibinfo{pages}{1085--1087}.
\newblock
\showISSN{0164-0925}
\urldef\tempurl%
\url{https://doi.org/10.1145/267959.269973}
\showDOI{\tempurl}


\bibitem[\protect\citeauthoryear{Zhang, Askarov, and Myers}{Zhang
  et~al\mbox{.}}{2011a}]%
        {Zhang-CCS11}
\bibfield{author}{\bibinfo{person}{Danfeng Zhang}, \bibinfo{person}{Aslan
  Askarov}, {and} \bibinfo{person}{Andrew~C. Myers}.}
  \bibinfo{year}{2011}\natexlab{a}.
\newblock \showarticletitle{Predictive Mitigation of Timing Channels in
  Interactive Systems}. In \bibinfo{booktitle}{\emph{Proceedings of the 18th
  ACM Conference on Computer and Communications Security}}
  \emph{(\bibinfo{series}{CCS '11})}. \bibinfo{publisher}{Association for
  Computing Machinery}, \bibinfo{pages}{563--574}.
\newblock
\showISBNx{9781450309486}
\urldef\tempurl%
\url{https://doi.org/10.1145/2046707.2046772}
\showDOI{\tempurl}


\bibitem[\protect\citeauthoryear{Zhang, Askarov, and Myers}{Zhang
  et~al\mbox{.}}{2012}]%
        {Zhang-PLDI12}
\bibfield{author}{\bibinfo{person}{Danfeng Zhang}, \bibinfo{person}{Aslan
  Askarov}, {and} \bibinfo{person}{Andrew~C. Myers}.}
  \bibinfo{year}{2012}\natexlab{}.
\newblock \showarticletitle{Language-Based Control and Mitigation of Timing
  Channels}. In \bibinfo{booktitle}{\emph{Proceedings of the 33rd ACM SIGPLAN
  Conference on Programming Language Design and Implementation}} (Beijing,
  China) \emph{(\bibinfo{series}{PLDI '12})}. \bibinfo{publisher}{Association
  for Computing Machinery}, \bibinfo{address}{New York, NY, USA},
  \bibinfo{pages}{99--110}.
\newblock
\showISBNx{9781450312059}
\urldef\tempurl%
\url{https://doi.org/10.1145/2254064.2254078}
\showDOI{\tempurl}


\bibitem[\protect\citeauthoryear{Zhang, Juels, Oprea, and Reiter}{Zhang
  et~al\mbox{.}}{2011b}]%
        {zhang-sp2011}
\bibfield{author}{\bibinfo{person}{Yinqian Zhang}, \bibinfo{person}{Ari Juels},
  \bibinfo{person}{Alina Oprea}, {and} \bibinfo{person}{Michael~K Reiter}.}
  \bibinfo{year}{2011}\natexlab{b}.
\newblock \showarticletitle{Homealone: Co-residency detection in the cloud via
  side-channel analysis}. In \bibinfo{booktitle}{\emph{2011 IEEE symposium on
  security and privacy}}. IEEE, \bibinfo{pages}{313--328}.
\newblock


\bibitem[\protect\citeauthoryear{Zhang and Reiter}{Zhang and Reiter}{2013}]%
        {zhang-ccs2013}
\bibfield{author}{\bibinfo{person}{Yinqian Zhang} {and}
  \bibinfo{person}{Michael~K. Reiter}.} \bibinfo{year}{2013}\natexlab{}.
\newblock \showarticletitle{D\"{u}Ppel: Retrofitting Commodity Operating
  Systems to Mitigate Cache Side Channels in the Cloud}. In
  \bibinfo{booktitle}{\emph{Proceedings of the 2013 ACM SIGSAC Conference on
  Computer \& Communications Security}} \emph{(\bibinfo{series}{CCS '13})}.
  \bibinfo{publisher}{Association for Computing Machinery},
  \bibinfo{pages}{827--838}.
\newblock
\showISBNx{9781450324779}
\urldef\tempurl%
\url{https://doi.org/10.1145/2508859.2516741}
\showDOI{\tempurl}


\end{thebibliography}

\appendix


\section{Decoy-path side channels}
\label{apx:example}

We use the following snippet to show how existing constant-time protection solutions struggle to maintain both memory safety and real execution invariants along decoy paths, ultimately introducing new side channels for attackers to detect decoy paths.

{\centering
\begin{lstlisting}[language=C, style=inlineCode,numbers=left]
char last_result;
char tableA[8192];
char tableB[4096];

char secret_hash(unsigned int secret) {
  if (secret < 4096) {
    register char tmp = tableB[secret];
    tmp ^= tableA[secret];
    last_result = tmp;
  }
  return last_result;
}
\end{lstlisting}
\par}

The {\tt if} condition at line 6 guards the statement at lines 7-9 (two read operations followed by one write operation). Let us consider the case $4096\mathrel{<=}\textsf{secret}\mathrel{<}8192$. All the state-of-the-art solutions~\cite{Coppens-SP09,Wu-ISSTA18,RACCOON-SEC15,Soares-CGO21} would also run the corresponding decoy path (statements inside the condition, normally executed only when $\textsf{secret}\mathrel{<}4096$), but with different code transformations. The approach of Coppens et al.~\cite{Coppens-SP09} rewires the memory accesses at lines 7-9 to touch a shadow address, therefore allowing an attacker to detect decoy path execution by observing (three) accesses to the shadow address. SC-Eliminator~\cite{Wu-ISSTA18} preloads both tables before executing the branch, but executes the read/write operations at lines 7-9 with unmodified addresses, introducing a decoy out-of-bounds read at line 7. Such memory safety violation might cause an exception if the memory after $\textsf{tableB}$ is unmapped, which, since the exception is left unmasked, would terminate the program and introduce a termination-based decoy-path side channel. Raccoon~\cite{RACCOON-SEC15} closes such termination side channels by masking the exception, but this strategy also introduces an exception handling-based decoy-path side channel. The approach of Soares et al.~\cite{Soares-CGO21}, on the other hand, replaces such an unsafe read access with an access to a shadow address, which however introduces the same decoy-path side channel discussed for Coppens et al.~\cite{Coppens-SP09}.

Finally, even assuming no exception is caused by the out-of-bounds read at line 7 and that we can even eliminate the out-of-bounds behavior altogether without introducing other side channels, an attacker can still trivially detect decoy-path execution by side channeling the read at line 8. The shadow access of Coppens et al.~\cite{Coppens-SP09} would leak decoy-path execution as discussed, but so will all the other solutions~\cite{RACCOON-SEC15,Soares-CGO21,Wu-ISSTA18}, which would allow an in-bound access at offset $4096\mathrel{<=}\textsf{secret}\mathrel{<}8192$ to $\textsf{tableA}$. Such access would never happen during real execution, breaking a program invariant on a decoy path and introducing a decoy data-flow side channel an attacker can use to again detect decoy-path execution.

In contrast, \proj's combined CFL/DFL strategy would instead ensure the very same data accesses during real or decoy execution, preserving program invariants and eliminating decoy-path side channels by construction. \cready{Table~\ref{tab:comparison} provides a detailed comparison between \proj and prior solutions.}

\begin{table*}[t]
\begin{footnotesize}
\centering
\caption{Technical, security, and compatibility features from state-of-the-art solutions vs. \proj.\label{tab:comparison}}
\vspace{-1.25em}
\adjustbox{max width=0.95\linewidth}{
\begin{tabular}{|l|c|c|c|c|c|c|}
\hline
\multicolumn{1}{|c|}{\textit{Feature}} & \textbf{Coppens et al.} & \textbf{Raccoon} & \textbf{SC-Eliminator} & \textbf{Soares et al.} & \textbf{\proj} \\
\hline
control flows & predicated & transactional & hybrid & hybrid & linearization \\
data flows & - & Path ORAM & preloading & - & linearization \\ 
loop handling strategy & unroll  & unroll & unroll & unroll & just-in-time \\ 
integration with compiler & backend & IR level & IR level & IR level & IR level \\
sensitive region identification & user annotations & annot. + static analysis & annot. + static analysis & user annotations & profiling (taint) \\
\hline
decoy-path side channels & shadow accesses & read/write accesses & read/write accesses & shadow, safe read/write accesses & no \\
fix variable-latency instructions (e.g., {\tt div}) & no & sw emulation & no & no & sw emulation \\
threat model & code & code+data & code+data* & code & code+data \\
\hline
variable-length loops & no & no & decoy paths till bound & no & yes \\
indirect calls & no & - & - & - & yes \\
recursion & fixed-depth** & fixed-depth & - & - & yes \\
spatial safety preserved & yes & no & no & yes & yes \\
supported data pointers & - & arrays & arrays & arrays & no restrictions \\
\hline
\end{tabular}}
\label{tab:features}\\
\vspace{0.25em}
{\hspace{1.2em}\scriptsize**~unimplemented}
\hspace{11.6em}{\scriptsize*~cache line with preloading}\hspace{4.2em}
\end{footnotesize}
\vspace{-1mm}
\end{table*}

\section{Conditional Selection}
\label{apx:select}

The {\tt ct\_select} primitive of \textsection\ref{se:cfl} can be instantiated in different ways. We studied how the LLVM compiler optimizes different schemes for constant-time conditional selection to pick the best possible alternative(s) in \proj.

For the discussion we consider the pointer selection primitive that we use to differentiate decoy and real store operations (i.e., to conditionally select whether we should actually modify memory contents), evaluating the alternatives listed \cready{below:}

\vspace{1mm}
{\centering
\adjustbox{max width=0.95\columnwidth}{
\begin{tabular}{| l | l | c | c |} 
\hline
Scheme & C equivalent & \taken values & CFL overhead \\ 
\hline\hline
1 & {\tt asm cmov} & {\tt\{0;1\}} & 9.6x \\ 
2 & {\tt ptr = taken ? ptr : ({\bf void}*)NULL } & {\tt\{0;1\}} & 7.5x \\
3 & {\tt ptr = ({\bf void}*)((size\_t)ptr \& (-taken))} & {\tt(size\_t) \{0;1\}} & 7.3x \\
4 & {\tt ptr = ({\bf void}*)((size\_t)ptr \& taken)} & {\tt\{0;0xff..ff\}} & 7.7x \\ 
5 & {\tt ptr = ({\bf void}*)((size\_t)ptr * taken) } & {\tt\{0;1\}} & 7.2x \\
\hline
\end{tabular}
}
\par}

\vspace{2mm}
We assume to operate on a {\tt void* ptr} pointer given as input to the {\tt ct\_store} primitive, and \taken values being $1$ on real paths and $0$ on decoy ones unless otherwise stated. Instead of reporting end-to-end overheads, we mask the slowdown from DFL by configuring CFL to use a single shadow variable as in the solution of Coppens et al.~\cite{Coppens-SP09}, then we compute the relative slowdowns of our protected {\tt mulmod} wolfSSL version (\textsection\ref{se:case-study}) for the different {\tt ct\_select} schemes, using the default non-CT implementation (W=1) as baseline.

Scheme 1 forces the backend to emit {\tt cmov} instructions at the assembly level, similarly to predicated execution mechanism we discussed in \textsection\ref{se:background}. As this choice constrains the optimizer's decisions, it turns out to be the worst performing alternative just as expected.
Scheme 2 is essentially an LLVM IR {\tt select} statement around the \taken indirection predicate, on which the compiler can reason about and optimize, then after IR-level optimizations the backend for most of its occurrences emits a {\tt cmov} instruction as in scheme 1, testing the value of the \taken variable for conditional assignment. 

Thus, we investigated different alternatives that could avoid relying on condition flags that, besides, get clobbered in the process and may require frequent recomputation.
While mask-based schemes 3 and 4 seemed at first the most promising avenues, it turned out that the additional operation needed either to generate the \taken mask from a boolean condition (scheme 3), or to maintain it at run-time and combining it with the boolean conditions coming from branch decisions (scheme 4), made these schemes suboptimal.
Scheme 5 resulted in the most simple and most efficient one between the solutions we tested, producing the lowest overhead as it unleashes several arithmetic optimizations (e.g. peephole, global value numbering) at IR and backend optimization levels.

\section{Striding}
\label{apx:striding}

One of the key performance enablers that back our radical approach is the ability to stride over object fields efficiently. We thoroughly tested different possible implementations, and designed different solutions based on the size of the field that should be strode and the granularity $\lambda$ at which the attacker could observe memory accesses. Several of these solutions leverage CPU SIMD Extensions: while we focused on AVX2 and AVX512 instructions for x86 architectures, the approach can easily be extended to other architectures supporting vectorization extensions (e.g., ARM SVE~\cite{ARMSVE-MICRO17} , RISC-V ``V''~\cite{RISCV-V}).
We group our solutions in three categories: simple object striding, vector {\tt gather}/{\tt scatter} operations, and vector bulk {\tt load}/{\tt stores}.

\paragraph{Simple Object Striding}
Given an access on pointer {\tt ptr} over some field {\tt f}, the most simple solution is to just access linearly {\tt f} at every $\lambda$-th location. We perform each access using a striding pointer {\tt s\_ptr} at the offset $\texttt{ptr}~\%~\lambda$ of the $\lambda$-th location, so that while striding a field {\tt s\_ptr} will match the target pointer {\tt ptr} exactly once, on the location where the memory access should happen. For load operations we conditionally maintain the value loaded from memory, propagating only the real result over the striding, while for store operation we load every value we access, conditionally updating it at the right location (\textsection\ref{ss:dfl-wrappers}). We report a simplified\footnote{Additional, constant-time logic is present in the implementation to deal with corner cases, so to avoid striding outside the object if not aligned correctly in memory.} snippet of a striding load operation for a {\tt uint8\_t}, where the conditional assignment is eventually realized e.g. using a {\tt cmov} operation:

{\centering
\begin{lstlisting}[language=C, style=inlineCode]
uint8_t ct_load(uint8_t* field, uint8_t* field_end, uint8_t* ptr) {
  // Default result
  uint8_t res = 0;

  // Get the offset of the pointer with respect to LAMBDA
  uint64_t target_lambda_off = ((uint64_t) ptr) & (LAMBDA-1);

  // Iterate over the possible offsets
  for(uint8_t* s_ptr = field; s_ptr < field_end; s_ptr += LAMBDA) {
    // Compute the current ptr
    uint8_t* curr_ptr = s_ptr + target_lambda_off;

    // Always load the value
    uint8_t _res = *(volatile uint8_t*)curr_ptr;

    // If curr_ptr matches ptr, select the value
    res = (curr_ptr == ptr)? _res : res;
  }
    return res;
}
\end{lstlisting}
\par}

\vspace{-1mm}
\paragraph{Vector gather/scatter Operations}
While for small fields the simple striding strategy presented above performs relatively well, larger sizes offer substantial room for improvement by leveraging SIMD extensions of commodity processors. Vectorization extensions offer instructions to gather (or scatter) multiple values in parallel from memory with a single operation. This allows us to design striding algorithms that touch in parallel $N$ memory locations (up to 16 for x86 AVX512 on pointers that fit into 32 bits). 
Our algorithm maintains up to $N$ indexes in parallel, from which to access memory from $N$ different locations at the same time. We manage in parallel the multiple accessed values similarly to for simple object striding, with the $N$ results being merged with an horizontal operation on the vector to produce a single value for loads.

\vspace{-1mm}
\paragraph{Vector Bulk load/stores}
Vectorization extensions allow us to load multiple values from memory at once, but a {\tt gather}/{\tt scatter} operation is in general costly for the processor to deal with. Depending on the value of $\lambda$ (e.g., with $\lambda\mathrel{=}1$ or $\lambda\mathrel{=}4$) most of the values could lie on the same cache line. Therefore we further optimized DFL with a third option which simply uses SIMD extensions to access a whole cache line (or half of it with {\tt AVX2}) with a single operation, thus touching all the bytes in that line. In case of loads, the accessed vector gets conditionally propagated with constant-time operations based on the real address to be retrieved, while for stores it gets conditionally updated, and always written back. 

\vspace{-1mm}
\paragraph{Sizing}
Building on empirical measurements on Skylake X and Whiskey Lake microarchitectures, we came up with a simple decision procedure to choose the best possible handler by taking into account the $\textit{size}$ of the field to stride, and the granularity $\lambda$ at which should be strode. Table~\ref{tab:striding} reports the average number clock cycles we measured for a striding handler given different values of $\textit{size}$ and $\lambda$. We list only the values for load operations for brevity, as the store handler incur in similar effects. We computed the number of cycles needed for each handler to stride over an object as the average of 1000 executions of the same handler, each measured using {\tt rdtscp} instructions. The SIMD based measurements are based on AVX2.

\begin{table}[t!]
\begin{scriptsize}
\centering
\caption{Clock cycles for different load striding handlers.\label{tab:striding}}
\vspace{-1em}
\begin{tabular}{| l | c | c | c |} 
\hline
handler & $\lambda$ & size & cycles \\ 
\hline\hline
simple & 64 & 64 & 4 \\ 
gather & 64 & 64 & 17 \\ 
bulk   & 64 & 64 & 9 \\
\hline
simple & 64 & 512 & 17 \\ 
gather & 64 & 512 & 17 \\ 
bulk   & 64 & 512 & 26 \\ 
\hline
simple & 64 & 1024 & 34 \\ 
gather & 64 & 1024 & 25 \\ 
bulk   & 64 & 1024 & 44 \\ 
\hline
\hline
simple & 4 & 64 & 34 \\ 
gather & 4 & 64 & 22 \\ 
bulk   & 4 & 64 & 11 \\
\hline
simple & 4 & 512 & 298 \\ 
gather & 4 & 512 & 116 \\ 
bulk   & 4 & 512 & 44 \\ 
\hline
simple & 4 & 1024 & 586 \\ 
gather & 4 & 1024 & 226 \\ 
bulk   & 4 & 1024 & 101 \\ 
\hline
\end{tabular}
\end{scriptsize}
\end{table}

We can immediately notice how bulk loads are the most effective for small $\lambda$ values, as they allow for accessing whole cache lines in a single instruction, so this is the default choice for such values. 
The situation is more complex for higher $\lambda$ values. We speculate that the AVX set-up operation for the CPU, and for the management of parallel values which should be merged together to produce the result, are too expensive to be amortized by the few iterations required to stride small objects. Therefore in this case we choose the simple object striding strategy. However, for bigger objects the gather operation is the clear winner, resulting in the minimum overhead. The decision algorithm in pseudocode reads as:

{\centering
\begin{lstlisting}[language=C, style=inlineCode]
if (LAMBDA < 16) select bulk
else if ((striding_size / LAMBDA) < 8) select simple
else select gather
\end{lstlisting}
\par}

\section{Field Sensitivity}
\label{apx:field-heuristics}

We improved the field-sensitivity of the Andersen \cready{points-to} analysis of SVF in order to delay demotion to field-insensitivity and recover, partially or to a full extent, the intended object portions thanks to heuristic inspired by duck typing from programming language research. In short, our extension restricts the surface of the abstract object that can be dereferenced to further improve the field information precision. The extension is semantically sound for the programs we consider, and in most cases (90\% for wolfSSL) could refine the SVF results up to the single desired field. We describe our extension using the following running example:

{\centering
\begin{lstlisting}[language=C, style=inlineCode]
%struct.fp_int = type { i32, i32, [136 x i64] } ; size = 1096
%struct.ecc_point = type { [1 x %struct.fp_int], [1 x %struct.fp_int],
                               [1 x %struct.fp_int] } ; size = 3288
%struct.ecc_key = type { i32, i32, i32, i32, %struct.ecc_set_type*,
                  i8*, %struct.ecc_point, %struct.fp_int }; size = 4416
\end{lstlisting}
\par}

These datatype declarations in LLVM IR describe the {\tt fp\_int}, {\tt ecc\_point}, and {\tt ecc\_key} structures of wolfSSL. An expression {\tt i\{8, 32, 64\}} denotes an integer type of the desired bit width. LLVM IR uses pointer expressions for load and store operation that come from a GEP ({\tt GetElementPtr}) instruction such as {\tt \%p} below:

{\centering
\begin{lstlisting}[language=C, style=inlineCode]
%v = <some %struct.fp_int object>
%p = getelementptr %struct.fp_int, %struct.fp_int %v, %i32 0, %i32 1
\end{lstlisting}
\par}

The syntax of a GEP instruction is as follows. The first argument is the type, the second is the base address for the computation, and the subsequent ones are indices for operating on the elements of aggregate types (i.e. structures or arrays). The first index operates on the base address pointer, and any subsequent index would operate on the pointed-to expressions. Here {\tt \%p} takes the address of the second {\tt i32} field of a {\tt fp\_int} structure. What happens with SVF is that it frequently reports a whole abstract object {\tt ecc\_key} in the points-to set for {\tt \%p}: this information if used as-is would require DFL to access all the 4416 object bytes during linearization.

Starting from this coarse-grained information, our technique identifies which portions of a large object could accommodate the pointer computation. In this simple example, we have that {\tt ecc\_key} can host one {\tt fp\_int} as outer member (last field), and three more through its {\tt ecc\_key} member (second-last). Hence we refine pointer metadata to set of each second {\tt i32} field from these objects, and only four 4-byte locations now require access during DFL. In general, we follow the flow of pointer value computations and determine object portions suitability for such dereferencing as in duck typing.


\section{Recursion and Thread Safety}
\label{apx:constructs}

Our implementation is lackluster in two respects that we could address with limited implementation effort, which we leave to future work as the programs we analyzed did not exercise them.

The first concerns handling recursive constructs. Direct recursion is straightforward: we may predict its maximal depth with profiling and apply the just-in-time linearization scheme seen for loops, padding recursive sequences with decoy calls for depths shorter than the prevision, using a global counter to track depths. For indirect recursion, we may start by identifying the functions involved in the sequence, as they would form a strongly connected component on the graph. Then we may apply standard compiler construction techniques, specifically the inlining approach of~\cite{scc-toplas97} to convert indirect recursion in direct recursion, and apply the just-in-time linearization scheme discussed above.


The second concerns multi-threading. As observed by the authors of Raccoon~\cite{RACCOON-SEC15}, programs must be free of data races for sensitive operations. The linearized code presently produced by \proj is not re-entrant because of stack variable promotion (\textsection\ref{sss:dfl-opts}) and for the global variable we use to expose the current taken predicate to called functions. On the code transformation side, an implementation extension may be to avoid such promotion, then use thread-local storage for the predicate, and update the doubly linked lists for allocation sites atomically. As for DFL handlers, we may implement DFL handlers either using locking mechanisms, or moving to more efficient lockless implementations using atomic operations or, for non-small involved sizes (\textsection\ref{apx:striding}), TSX transactions.

\begin{table*}[t]
\begin{footnotesize}
\centering
\caption{Run-time overhead analysis of different \proj configurations (i.e., $\lambda\mathrel{=}1$, AVX2). The numbers for SC-Eliminator and Soares et al. were obtained from executing the publicly available artifact evaluation material for their papers.\label{tab:allbench}}
\vspace{-1em}
\adjustbox{max width=0.87\linewidth}{
\begin{tabular}{| l | l | c | c | c | c | c | c | c |} 
\cline{2-9}
\multicolumn{1}{c|}{} & program & AVX512 ($\lambda\mathbin{=}1$) & AVX512 ($\lambda\mathbin{=}4$) & AVX512 ($\lambda\mathbin{=}64$) & AVX2 ($\lambda\mathbin{=}4$) & AVX2 ($\lambda\mathbin{=}64$) & SC-Eliminator & Soares et al. \\ 
\hline
\multirow{8}{*}{\textsc{chronos}} & aes & 1.13 & 1.13 & 1.08 & 1.22 & 1.08 & 1.11 & 1.02 \\ 
& des & 1.12 & 1.19 & 1.14 & 1.36 & 1.15 & 1.09 & 1.00 \\ 
& des3 & 1.49 & 1.49 & 1.36 & 1.86 & 1.37 & 1.12 & 1.02 \\ 
& anubis & 1.29 & 1.29 & 1.12 & 1.55 & 1.22 & 1.06 & 1.00 \\ 
& cast5 & 1.13 & 1.13 & 1.06 & 1.25 & 1.12 & 1.06 & 1.02 \\ 
& cast6 & 1.13 & 1.13 & 1.08 & 1.13 & 1.07 & 1.05 & 1.00 \\ 
& fcrypt & 1.04 & 1.04 & 1.03 & 1.06 & 1.01 & 1.05 & 1.00 \\ 
& khazad & 1.13 & 1.13 & 1.09 & 1.23 & 1.07 & 1.30 & 1.00 \\ 
\hline
\multirow{2}{*}{\textsc{s-cp}} & aes\_core & 1.12 & 1.12 & 1.06 & 1.01 & 1.05 & 1.06 & 1.04 \\ 
& cast-ssl & 1.23 & 1.23 & 1.10 & 1.38 & 1.15 & 1.17 & 1.01 \\
\hline 
\multirow{6}{*}{\textsc{botan}} & aes & 1.05 & 1.05 & 1.03 & 1.09 & 1.01 & 1.01 & - \\ 
& cast128 & 1.02 & 1.02 & 1.01 & 1.03 & 1.01 & 1.05 & - \\ 
& des & 1.01 & 1.01 & 1.01 & 1.01 & 1.01 & 1.08 & - \\ 
& kasumi & 1.01 & 1.01 & 1.01 & 1.03 & 1.01 & 1.01 & - \\ 
& seed & 1.02 & 1.02 & 1.01 & 1.03 & 1.01 & 1.03 & - \\ 
& twofish & 1.14 & 1.14 & 1.12 & 1.21 & 1.15 & 1.04 & - \\
\hline
\multirow{3}{*}{\textsc{app-cr}} & 3way & 1.00 & 1.00 & 1.00 & 1.00 & 1.00 & 1.03 & 1.15 \\ 
& des & 1.24 & 1.24 & 1.09 & 1.27 & 1.06 & 1.08 & 1.11 \\ 
& loki91 & 1.51 & 1.51 & 1.43 & 1.48 & 1.48 & 1.97 & 1.24 \\
\hline
\multirow{4}{*}{\textsc{libgcrypt}} & camellia & 1.02 & 1.02 & 1.01 & 1.02 & 1.01 & 1.08 & 1.01 \\ 
& des & 1.06 & 1.06 & 1.06 & 1.05 & 1.09 & 1.03 & 1.01 \\ 
& seed & 1.18 & 1.18 & 1.10 & 1.21 & 1.01 & 1.15 & 1.01 \\ 
& twofish & 1.97 & 1.97 & 1.92 & 2.45 & 2.10 & 1.41 & 1.24 \\
\hline
\multirow{6}{*}{\textsc{raccoon}} & binsearch & 1.33 & 1.33 & 1.18 & 1.30 & 1.16 & - & - \\ 
& dijkstra & 3.87 & 3.45 & 1.51 & 2.83 & 1.50 & - & - \\ 
& findmax & 1.00 & 1.00 & 1.00 & 1.00 & 1.00 & - & - \\ 
& histogram & 2.66 & 2.66 & 1.68 & 4.39 & 1.87 & - & - \\ 
& matmul & 1.00 & 1.00 & 1.00 & 1.00 & 1.00 & - & - \\ 
& rsort & 1.87 & 1.87 & 1.84 & 1.50 & 1.45 & - & - \\ 
\hline
\multirow{5}{*}{\textsc{pycrypto}}
& aes              & 1.13 & 1.13 & 1.06 & 1.33 & 1.11 & - & - \\
& arc4             & 1.07 & 1.07 & 1.03 & 1.08 & 1.03 & - & - \\
& blowfish         & 5.07 & 5.07 & 3.17 & 10.58 & 3.23 & - & - \\
& cast             & 1.09 & 1.09 & 1.04 & 1.16 & 1.08 & - & - \\
& des3             & 1.06 & 1.06 & 1.04 & 1.08 & 1.05 & - & - \\
\hline
\multirow{3}{*}{\textsc{B/Rel}}
& tls-rempad-luk13 & 1.01 & 1.01 & 1.01 & 1.01 & 1.01 & - & - \\
& aes\_big         & 1.02 & 1.01 & 1.01 & 1.02 & 1.01 & - & - \\
& des\_tab         & 1.04 & 1.04 & 1.02 & 1.07 & 1.03 & - & - \\
\cline{1-9}
\multicolumn{1}{c|}{} & geomean (total) & 1.26 & 1.26 & 1.16 & 1.34 & 1.17 & 1.12 & 1.05 \\
\cline{2-9}
\end{tabular}
}
\vspace{0.5em}
\end{footnotesize}
\end{table*}

\section{Correctness}
\label{apx:correctness}

We discuss an informal proof of correctness of the \proj approach. As we anticipated in \textsection\ref{se:security}, to prove that our programs are semantically equivalent to their original representations, we break the claim into two parts: control-flow correctness and data-flow correctness For each part we assume that the other holds, so that the initial claim can hold by construction.

For control flows, we need to show that along real paths the transformed program performs all and only the computations that the original one would make. First, we rule out divergences from exceptional control flow since the original program is error-free and CFL sanitizes sequences that may throw (e.g., division) when in dummy execution. We then observe that by construction CFL forces the program to explore both outcomes of every branch, and the decision whether to treat each direction in dummy execution depends on the \taken predicate value. CFL builds this predicate as the combination of the control-flow decisions that the (original) program takes on the program state, and from the data flow argument decoy paths have no effects on such state. All the linearized branch directions will be executed as many times and in the same interleaving observable in the original program; as for the special loop case, the amount of real and decoy iterations depends on the original loop condition and the \taken predicate, so its real iterations closely match the original loop. This completes the argument. 


For data flows, we need to show that values computed in dummy execution cannot flow into real paths, and that decoy paths preserve memory safety. The points-to metadata fed to DFL load and store wrappers make the program access the same memory objects along both real and decoy paths, and allocation metadata ensure that those objects are valid: memory safety is guaranteed. Also, only real paths can change memory contents during a store, hence only values written by real paths can affect data loads. Thus, we only need to reason about data flows from local variables assigned in dummy execution. LLVM IR hosts such variables in SSA virtual registers, and at any program point only one variable instance can be live~\cite{Rosen-POPL88}. For a top-level linearized branch, a {\tt ct\_select} statement chooses the incoming value from the real path (\textsection\ref{ss:dummy}). For a nested branch (Figure~\ref{fig:dummy}) both directions may be part of dummy execution. Regardless of which value the inner {\tt ct\_select} will choose, the outer one eventually picks the value coming from the real path that did not contain the branch. Extending the argument to three or more nested branches is analogous: for a variable that outlives a linearized region, whenever such variable is later accessed on a real path, the value from a real path would assign to it (otherwise the original program would be reading an undefined value or control-flow correctness would be violated), while in decoy paths bogus value can freely flow. We discussed correctness for loops in \textsection\ref{ss:loops}.

\balance
\section{Complete Run-Time Overhead Data}
\label{apx:results}
Table~\ref{tab:allbench} shows the complete set of our performance-oriented experiments: we benchmarked \proj with different $\lambda$ values ($1, 4, 64$) and SIMD capabilities (AVX2 and AVX512), and also ran the artifacts from~\cite{Wu-ISSTA18} and~\cite{Soares-CGO21} on the same setup used for \textsection\ref{se:performance}. For the latter we did not try the Raccoon microbenchmarks, mostly due to compatibility problems and limitations of the artifacts, while the SCE suite was part of their original evaluation (Soares et al. leave the {\tt botan} group out of the artifact evaluation harness). Both systems provide much weaker security properties than \proj, yet the average overhead numbers we observe are similar. Also, the availability of AVX512 instructions brings benefits for the $\lambda\mathrel{=}64$ setting as they allow DFL to touch more cache lines at once over large object portions (\textsection\ref{apx:striding}). Interesting, protection for the presently unrealistic $\lambda\mathrel{=}1$ attack vector leads to overheads that are identical to the $\lambda\mathrel{=}4$ configuration for MemJam-like attacks, with {\tt  dijkstra} being the only exception (3.87x vs 3.45x).

\end{document}